\documentclass[10pt onecolumn,english,prd,showpacs,nofootinbib]{revtex4}
\usepackage[latin9]{inputenc}
\usepackage{graphicx}
\usepackage{amsmath}
\usepackage{amssymb}
\usepackage{amsfonts}
\usepackage{lscape}
\usepackage{url}
\usepackage{epsfig}
\usepackage{color}
\usepackage{bm}
\usepackage{babel}

\providecommand{\tabularnewline}{\\}

\providecommand{\tabularnewline}{\\}



\newcommand{\lzr}{{\left(z\right)}}
\newcommand{\nn}{\nonumber}

\def\be{\begin{equation}}
\def\ee{\end{equation}}
\def\bea{\begin{eqnarray}}
\def\eea{\end{eqnarray}}

\begin{document}
\title{Curvature vs Distances: testing the FLRW cosmology}

\author{Domenico Sapone}
\email{domenico.sapone@uam.es}

\author{Elisabetta Majerotto}
\email{elisabetta.majerotto@uam.es}

\author{Savvas Nesseris}
\email{savvas.nesseris@uam.es}

\affiliation{Departamento de F\'isica Te\'orica and Instituto de F\'isica Te\'orica, \\
Universidad Aut\'onoma de Madrid IFT-UAM/CSIC,\\
$28049$ Cantoblanco, Madrid, Spain}


\begin{abstract}
We test the FLRW cosmology by reconstructing in a model-independent way both the Hubble parameter $H(z)$ and the comoving distance $D(z)$ via the most recent Hubble and Supernovae Ia data. In particular we use: data binning with direct error propagation, the principal component analysis, the genetic algorithms and the Pad\'e approximation. Using our  reconstructions we evaluate the Clarkson {\it et al} test known as $\Omega_K(z)$, whose value is constant in redshift for the standard cosmological model, but deviates elsewise. We find good agreement with the expected values of the standard cosmological model within the experimental errors. Finally, we provide forecasts, exploiting the Baryon Acoustic Oscillations measurements from the Euclid survey.
\end{abstract}
\maketitle

\section{Introduction}

The observed accelerated expansion of the late-time universe, as evidenced by
a number of cosmological data like type Ia supernovae (SnIa) \cite{sn1a}, the cosmic
microwave background radiation \cite{cmb} (CMB) and large scale structure \cite{lss} (LSS)
came as a great surprise to cosmologists. It is quite straightforward to explain the effect within
the framework of Friedmann-Lema\^itre-Robertson-Walker (FLRW) cosmology by simply introducing
a cosmological constant or a more general (dynamical) dark energy component.
However, all such components give rise to severe coincidence and fine-tuning problems.
An alternative approach postulates that General Relativity is only accurate on small scales and
modifications at larger scales are needed. These modification would lead to the observed
late-time acceleration \cite{CD,BDEL,RM,fR}.

Recently, several works have tried to take a different approach, namely to study the effect of
large scale structure on the observed luminosity-to-redshift behavior of SnIa
(the first observable that has led to the conclusion that the Universe is accelerating).
The assumption that the Universe is homogeneous and isotropic is known to be violated
at late times. However, the scale at which the assumption
breaks down is still in debate, \cite{campusano}.
Nonetheless, recent studies have been able to reproduce the luminosity-to-redshift relation provided
that we live in a large region, called empty void, in which the matter density is less
than the spatial average density on large cosmological scales (\cite{LTB}, for a
recent review see \cite{voids-review}).
However, also these models are still in debate since they require the observer to be
located at the center of the void or at most within about few percent of the void scale radius \cite{BM},
thus disfavoring the model from a Copernican Principle (CP) point of view.

Despite all the theoretical efforts to understand the accelerated expansion of the Universe, see \cite{ks, sapone},
right now there is no model capable of providing singlehandedly a satisfactory explanation.
An alternative is to create null tests for the current paradigm and see if and how it breaks down.
For example, Clarkson {\it et al.} \cite{cbt} presented a new test for the CP which relies on a
consistency relation that exists within the homogenous and isotropic FLRW model between
the luminosity and the Hubble parameter. This relation is expected to hold exactly at all redshifts $z$,
and any deviation will point to a departure from the FLRW model. Indeed, to be precise, the test of Clarkson {\it et al.} tests deviations from a FLRW metric and not strictly from the CP, as pointed out by \cite{wil2, wil, s1}.

The existence of this consistency relation implies that if we have two separate experiments measuring
independently the comoving distance $D(z)$ (or equivalently the angular diameter distance or the luminosity distance)
and the Hubble parameter $H(z)$, then we are able to reconstruct the curvature parameter $\Omega_{K}$
at each redshift in a model-independent way. In a FLRW universe,
the curvature parameter should not depend on redshift, hence $\Omega_{K}$ should not vary,
so if we measure a variation of the curvature parameter over redshift,
this means that the assumption of homogeneity at large scales has to be rejected.

Since it was proposed, this test has raised interest in the astrophysics community.
Ref. \cite{avg, sha} have applied it by using available data; ref. \cite{s4} has discussed the deviation from the FLRW relation due to backreaction; refs. \cite{wil, feb, s2, s3} have predicted
the outcome of the test, i.e. the behavior of $\Omega_K$ with redshift, in the case of their particular
inhomogeneous cosmology or void model; ref. \cite{lar} has made predictions on future results
from the test if a toy-model of backreaction is assumed to be the correct cosmology;
ref. \cite{mor} has used it to measure the curvature of the Universe in a model-independent way. Also, many other tests of the standard model of cosmology have been outlined (for most recent work, see \cite{s5}, and for a review see \cite{Clarkson-review, Marra-review})

In particular, ref. \cite{avg} has used $H(z)$ data from passively evolving galaxies \cite{Jimenez:2003iv, Simon:2004tf} and
the SnIa of \cite{Kowalski:2008ez} and has found no indication of a deviation from FLRW.
Ref. \cite{mor} has used SnIa data of \cite{Amanullah:2010vv},
H(z) data from passively evolving galaxies \cite{Stern:2009ep} and baryonic acoustic oscillation (BAO)
data \cite{Gaztanaga:2008xz}, and the $H_0$ measurement of \cite{Riess:2009pu}.
Similarly, ref. \cite{sha} has used the SnIa of \cite{Hicken:2009dk} and the measurement
of $H_0$ of \cite{Riess:2009pu} or of \cite{Komatsu:2008hk}, in combination with $H(z)$ measurements
coming either from passively evolving galaxies \cite{Jimenez:2003iv, Stern:2009ep},
or from BAO data \cite{Percival:2009xn}, together with cosmic microwave background data \cite{Wang:2007mza}.
No evidence for deviations from FLRW nor from flatness was found.

In this context, our work has two objectives, building on the work of \cite{sha} and \cite{lar}.
The first objective is to improve the measurement of $\Omega_K$ by comparing four different
measurement techniques based both on binnings of the data and on very efficient
model-independent reconstructions of functions, and by using the most recent data.
The second is to make new up-to-date forecasts for the $\Omega_K$ test for future LSS and SnIa data.

The paper is organized as follows: in Sec.~\ref{sec:background} we describe the general background
equations used for our analysis. In Sec.~\ref{sec:reconstruction_H_D} we describe four different
methods to reconstruct the Hubble parameter and the comoving distance from two different data sets.
In particular we show: the data binning and direct error propagation method,
the principal component analysis (PCA), the genetical algorithms (GA) and the Pad\'e approximation.
In each subsection we report the results. Sec.~\ref{sec:forecasts} is devoted to the
Fisher matrix forecasts on the errors for the $\Omega_{K}$ at different $z$ using the
Euclid galaxy redshift survey\footnote{http://www.euclid-ec.org/}
\cite{Cimatti:2009is, RedBook} and future SnIa surveys.

\section{Background equations}\label{sec:background}

Here we review the basic equations and notation for the background evolution.
The evolution of the dark energy can be expressed by the present dark energy density $\Omega_{DE}$
and by a time-varying equation of state:
\be
w(z)=\frac{p}{\rho}
\label{eq:parameterOfState}\,.
\ee
Where $p$ and $\rho$ are the pressure and energy density of dark energy, respectively.
The dark energy density equation is $\rho(z)=\rho(0)a^{-3(1+\hat{w})}$ and
\be
\hat{w}(z)=\frac{1}{\log(1+z)}\int_{0}^{z}\frac{w(z')}{1+z'}dz' \,.
\label{eq:densit}
\ee
The Hubble parameter $H(z)$ and the angular diameter distance $D_{A}(z)$,
in a flat universe with $\Omega_{m}+\Omega_{DE}=1$, are
\be
H^{2}\left(z\right)=H_{0}^{2}[\Omega_{m}(1+z)^{3}+(1-\Omega_{m})(1+z)^{3(1+\hat{w})}]
\label{eq:HubbleParam}
\ee
and
\be
D_{A}(z)=\frac{c}{1+z}\int_{0}^{z}\frac{dz}{H(z)}\,.
\label{eq:flatAngDist}
\ee
In a general FLRW model with curvature, the angular diameter distance can be written as:
\be
D_{A}(z)=\frac{c}{1+z}\frac{1}{H_{0}\sqrt{-\Omega_{K}}}\sin\left(\sqrt{-\Omega_{K}}\int_{0}^{z}{dz'\frac{H_{0}}{H(z')}}\right)\,.
\label{eq:curvAngDist}
\ee
where $\Omega_{K}$ is the curvature parameter today.

We can invert Eq.~(\ref{eq:curvAngDist}) to obtain an expression for the curvature parameter $\Omega_{K}$ that
depends on the Hubble parameter $H(z)$ and comoving distance $D(z)=(1+z)D_{A}(z)$, see \cite{cbt}:
\be
\Omega_{K}(z)=\frac{\left[H\left(z\right)D_{,z}\left(z\right)\right]^{2}-1}{\left[H_{0}D\left(z\right)\right]^{2}}\,,
\label{eq:omegak}
\ee
where the comma refers to the derivative with respect to the redshift.
The above equation tells us how we can measure the curvature parameter
in a model-independent way from Hubble rate and distance measurements.
If we live in a FLRW universe, then the curvature parameter is independent of redshift, i.e. $\Omega_{K}(z)$ should be a constant $\Omega_{K}(z)=\Omega_{K}$;
however, if we measure a variation of the curvature parameter, this would indicate that the homogeneity of the large-scale universe is violated.

In order to test the values of $\Omega_{K}$ at different redshifts we need to combine two independent measurements:
that of the Hubble parameter $H(z)$ and that of the comoving distance $D(z)$.
In the next section we will use the Hubble parameter data set from \cite{moresco_etal} and the SnIa magnitude data-set from the SCP ``Union2.1'' \cite{union, suzuki_etal} to reconstruct the curvature parameter as a function of redshift.

\section{Reconstructing $H(z)$ and $D(z)$}\label{sec:reconstruction_H_D}

In this section we present four different ways to reconstruct the curvature parameter using measurements of the Hubble parameter and comoving distance from SnIa.

\subsection{Data sets}\label{sec:data}

First, we briefly describe the data sets that we will use in the present analysis. These are:

\begin{itemize}
\item{The Hubble parameter data that directly probe $H(z)$, in the most recent compilation given by Moresco et al. \cite{moresco_etal}.
The authors implemented a {\em differential approach} to evaluate the Hubble parameter;
first they chose from different catalogs early-type galaxies and then they selected only the most massive,
red elliptical galaxies which are passively evolving and do not manifest any signature of ongoing star formation.
The Hubble parameter is then given by
\be
H(z) = -\frac{1}{1+z}\frac{{\rm d}z}{{\rm d}t}
\ee
where ${\rm d}z$ is simply given by the difference in redshift of two galaxies and ${\rm d}t$ is given by
their differential dating of star populations.
}

\item{The SnIa data that probe the luminosity distance $d_L(z)\equiv (1+z) D(z)$,
where $D(z)$ is the comoving distance. In particular, we use the ``Union2.1''
set of 580 SnIa of Suzuki et al. \cite{suzuki_etal}
\footnote{The SnIa data can be found in http://supernova.lbl.gov/Union/ and
in \cite{suzuki_etal}}. The data are given in terms of the distance modulus
\bea
\mu(z)&\equiv& m(z)-M=5 \log_{10}\left(d_L(z)\right)+25\\
&=&5 \log_{10}\left(H_0 d_L(z)\right)+\mu_0,
\label{eq:dist-mod}
\eea
where $m(z)$ is the apparent magnitude at peak brightness, $M$ is the absolute magnitude
and $\mu_0=42.38-5 \log_{10}h$ with $h=H_0/[100~\textrm{km}~\textrm{sec}^{-1}~ \textrm{Mpc}^{-1}]$.
The chi-square is then
\be
\chi^2_{SnIa}=\sum_{i=1}^{580}\left(\frac{\mu_{obs}(z_i)-\mu_{th}(z_i)}{\sigma_i}\right)^2.
\ee
For simplicity we only consider the case where the covariance matrix of the SnIa data is diagonal.}
\end{itemize}

\subsection{Binning SnIa and $H(z)$ data}\label{sec:binning}

\begin{figure}
\includegraphics[scale=0.68]{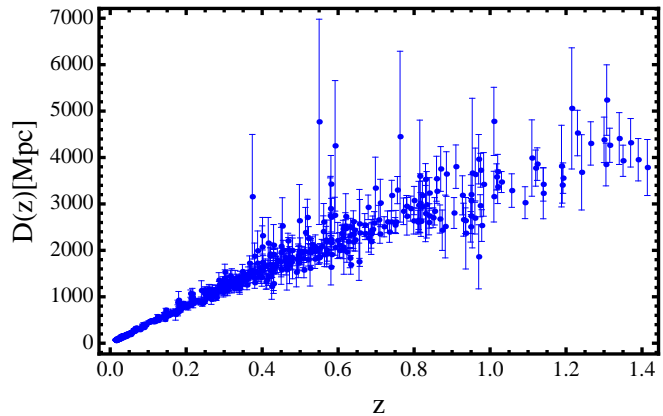}
\caption{Comoving distances computed directly from the 580 SnIa of the ``Union2.1'' compilation \cite{union} by using Eq. (\ref{eq:com-dist}).}
\label{fig:comoving-distances}
\end{figure}

The first technique to measure $\Omega_K(z)$ consists in evaluating it in several redshift bins by
directly computing the comoving distance $D(z)$ and its derivative $D_{,z}$
from the SnIa data and by using the $H(z)$ values measured from passively evolving galaxies data.
To compute $D(z)$, we simply invert  Eq. (\ref{eq:dist-mod}) that expresses
the  distance modulus $\mu$ as a function of $D(z)$:
\be
D(z) = \frac{10^{\frac{\mu -25}{5}}}{1+z}\,.
\label{eq:com-dist}
\ee
The resulting comoving distances computed from the ``Union2.1'' SnIa are shown in Fig. \ref{fig:comoving-distances}.
The obtained data are then divided into bins. We do this in two different ways, in order to understand which way produces the best results, and to control the dependence on the specific binning.

\subsubsection{First binning criterion} \label{sec:bins1}
\begin{figure}
\includegraphics[scale=0.68]{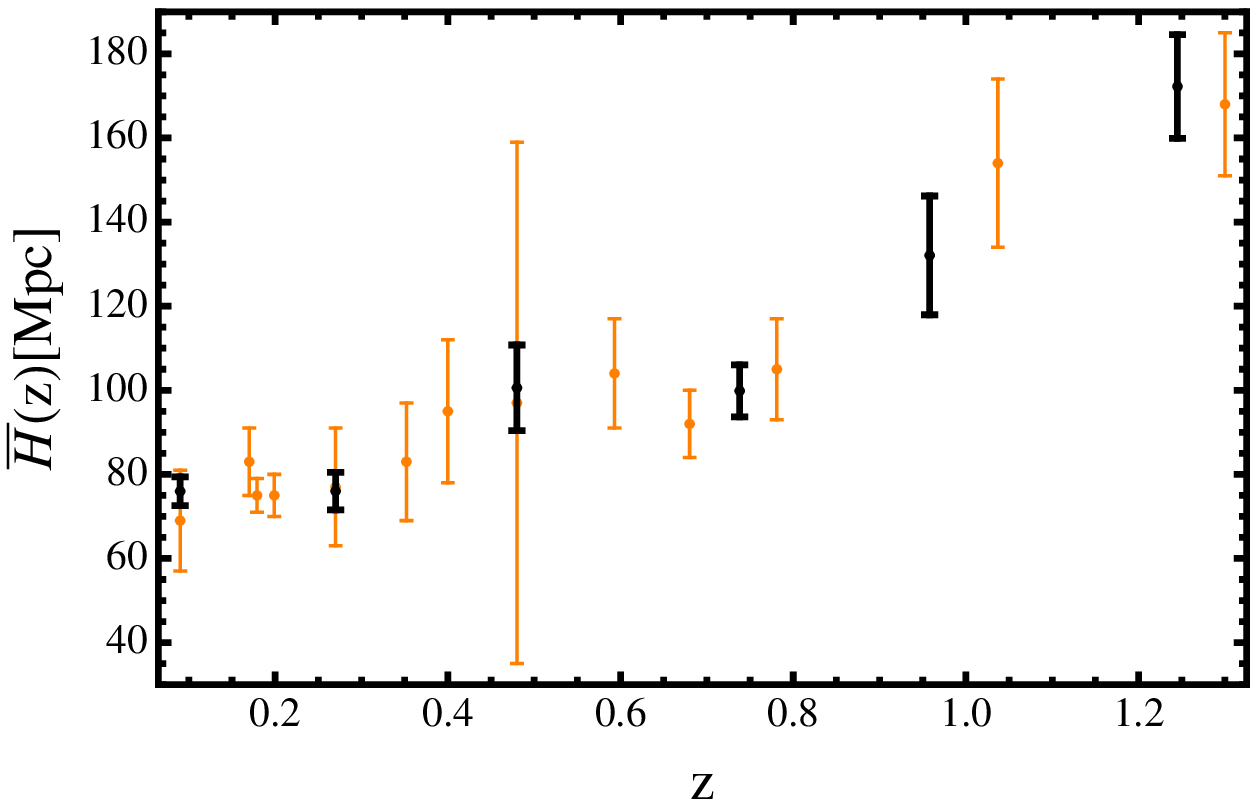}\\
\includegraphics[scale=0.68]{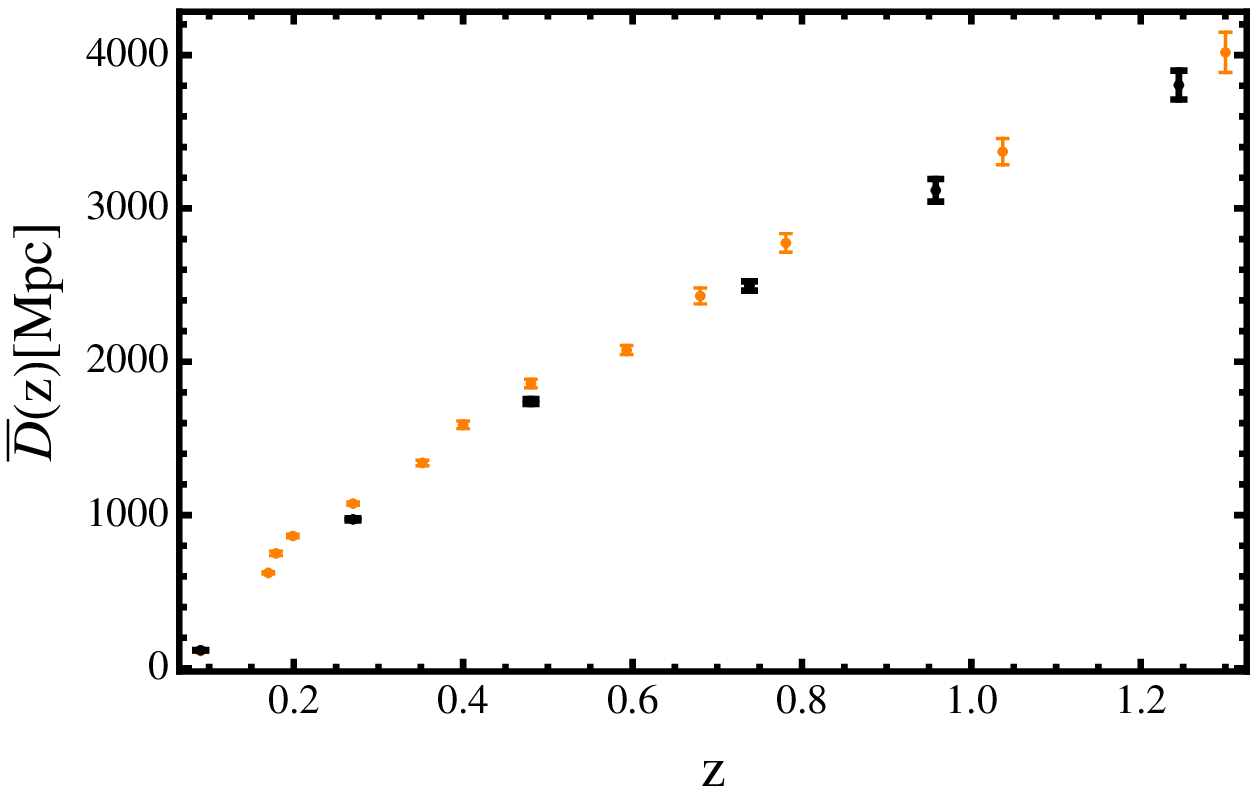}\quad
\includegraphics[scale=0.68]{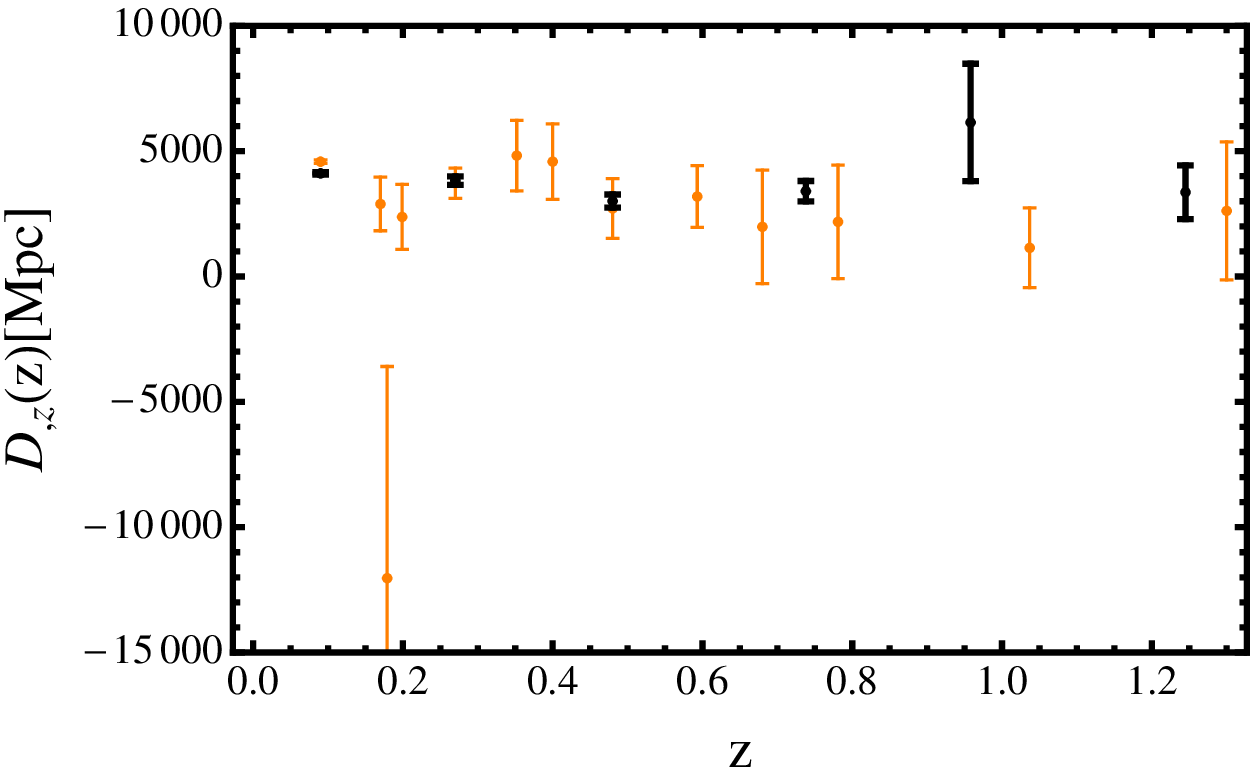}
\caption{The mean Hubble function $\bar H$, comoving distance $\bar D$, and derivative of the comoving distance $D_{,z}$ computed with the binning of Secs. \ref{sec:bins1} (orange thin error bars) and \ref{sec:bins2} (black thick error bars). We omit to show horizontal error bars in order not to make the plot too crowded and confusing.}
\label{fig:H-D-dD}
\end{figure}

In the first approach, we aim at maximising the number of bins. Since the number $n_H = 19$ of $H(z)$ data is quite small, we choose a number of bins as close as possible to $n_H$. To do so, we use the following procedure. We
split the redshift interval $\Delta z_{i,i+1}$ between each pair of $H(z)$ data, $H(z_i)$ and $H(z_{i+1})$,  into two equal parts. We take $z=0$ as the initial redshift of the first bin.
Since the SnIa's highest redshift is $z=1.414$, the last bin does not contain any SnIa, and we discard it. This means that we also have to discard the last two $H(z)$ data points. The resulting bins are given in Tab. \ref{tab:bins1}.
\begin{table}
\begin{centering}
\begin{tabular}{|c|c|c|c|c|c|c|c|c|c|}
\hline
\textbf{Bins}  & $z_{min}$ &$z_{max}$ & $z_H$ & $\bar z$ &$\#$ of SnIa & $\#$ of H(z) & $\bar D$ & $D_{,z}$ & $\Omega_{K}$\tabularnewline
\hline
$1$&$0.00$&$0.13$&$0.09$ &$ 0.044 $& $194 $& $1$ & $114.29 \pm 0.68$ & $4578\pm 71$ & $138\pm 490$\tabularnewline
\hline
$2$&$0.13$&$0.17$& $0.17$ &$0.16$&$20$& $1$ & $622.9\pm 7.7$ & $2892\pm 1068$ & $-15\pm 21$ \tabularnewline
\hline
$3$&$0.17$& $0.19$&$0.18$&$0.18$&$12$& $1$ & $750\pm 14$ & $-12036\pm 8445$ & $237\pm 375$\tabularnewline
\hline
$4$&$0.19$&$0.23$&$0.20$&$0.21$&$24$&  $1$ & $864\pm 11$ & $2375\pm 1296$ & $-14.3\pm 8.6$ \tabularnewline
\hline
$5$&$0.23$& $0.31$&$0.27$&$0.27$&$54$&  $1$ & $1075\pm 11$ & $3719\pm 601$ & $-1.3\pm 6.3$ \tabularnewline
\hline
$6$&$0.31$&$0.38$&$0.35$&$0.34$&$37$&  $1$ & $1340\pm 18$ & $4821\pm 1407$ & $7\pm 11$ \tabularnewline
\hline
$7$&$0.38$&$0.44$&$0.40$&$0.41$&$43$&  $1$ & $1589\pm 25$ & $4581\pm 1505$ & $7\pm 10 $\tabularnewline
\hline
$8$&$0.44$&$0.54$&$0.48$&$0.49$&$43$&  $1$ & $1858\pm 28$ & $2713\pm 1190$ & $-1.1\pm 5.7$ \tabularnewline
\hline
$9$&$0.54$&$0.64$&$0.59$&$0.59$&$45$&  $1$ & $2076\pm 29$ & $3189\pm 1233$ & $0.9\pm 3.8$\tabularnewline
\hline
$10$&$0.64$&$0.73$&$0.68$&$0.68$&$23$&  $1$ & $2429\pm 52$ & $1980\pm 2266$ & $-1.8\pm 2.4$\tabularnewline
\hline
$11$&$0.73$&$0.83$&$0.78$&$0.79$&$22$&  $1$ & $2775\pm 61$ & $2181\pm 2262$ & $-0.9\pm 2.6$ \tabularnewline
\hline
$12$&$0.97$&$1.17$&$1.04$&$1.04$& $19$&  $1$ & $3370\pm 85$ & $1143\pm 1591$ & $-1.0\pm 1.4$ \tabularnewline
\hline
$13$&$1.17$&$1.37$&$1.30$&$1.26$&$13$&  $1$ & $4018\pm 131$ & $2616\pm 2754$ & $1.2\pm 4.6$\tabularnewline
\hline
\end{tabular}
\par\end{centering}
\caption{Properties of bins described in Sec. \ref{sec:bins1} and values of the comoving distance $\bar D$, derivative of the comoving distance $D_{,z}$ and curvature parameter $\Omega_K$ measured in each bin. With $z_{min}$ and $z_{max}$ we indicate the left and right edge of each bin, respectively. $z_H$ refers to the redshift to which the mean distance $\bar D$ has been assigned, while $\bar z$ is the mean redshift of the bin.
\label{tab:bins1}}
\end{table}
In each bin we then compute the comoving distance $\bar{D}$ as the weighted mean of the different $D_i$s contained in the bin, weighted by the square of the error on $D_i$, $\sigma_i^2$, and assign $\bar{D}$ to the redshift $z_H$.
In principle, we should assign $\bar{D}$ to the mean redshift $\bar{z}$ of each bin, but, as can be seen from Table \ref{tab:bins1}, $\bar{z}$ and $z_H$ are very close: only for the first bin the difference is larger than 9\%, but here we will see that errors on $\Omega_K$ are extremely large anyway and constraints on $\Omega_K$ will not be useful in practice.
 The error on $\bar{D}$ is then
$\sigma_{\bar D} = 1/\sqrt{\sum_i 1/\sigma_i^2}$.
To compute the derivative $D_{,z}$ we use one of the formulas for its discrete approximation:
\be
D_{,z}(z) \simeq \frac{D(z+\Delta z_1)-D(z-\Delta z_2)}{\Delta z_1+\Delta z_2}\,.
\label{eq:finite-der}
\ee
In order to apply this formula to our data, we split each bin into two parts, taking $z_H$ as splitting point. We then compute $\bar D$ in each sub-bin, obtaining two mean distances $\bar D_{left}$ and $\bar D_{right}$. We assign $\bar D_{left}$ and $\bar D_{right}$ to the average redshift of the corresponding sub-bin, $\bar z_{left}$ and $\bar z_{right}$, and finally compute the (approximated) derivative as
\be
D_{,z}(z_H) = \frac{\bar D(\bar z_{right}) - \bar D(\bar z_{left})}{\bar z_{right} - \bar z_{left}}.
\label{eq:der-num}
\ee
The error on the derivatives is obtained via the simple propagation of errors formula.
We find that not all sub-bins contain at least one SnIa, so we eliminate from the sample all bins which have no SnIa in either the left or the right corresponding sub-bin. This leaves us with only 13 bins; as a consequence, only 13 out of 19 $H(z)$ data points and only 549 out of 580 SnIa data points are used.
In Fig. \ref{fig:H-D-dD} we show (orange thin error bars) $\bar H$, $\bar D$ and $D_{,z}$ obtained with this binning.

We then compute $\Omega_K(z_H)$ at each redshift $z_H$ by using Eq. (\ref{eq:omegak}), where, as already said, $D(z_H)$ and $D_{,z}(z_H)$ are computed from SnIa data as explained above, $H(z_H)$ come from passively evolving galaxies data and where we use  $H_0 = 73.8 \pm 2.4$, as measured from the Hubble Space Telescope (HST) and the Wide Field Camera 3 \cite{HST}.

\begin{figure}
\includegraphics[scale=0.68]{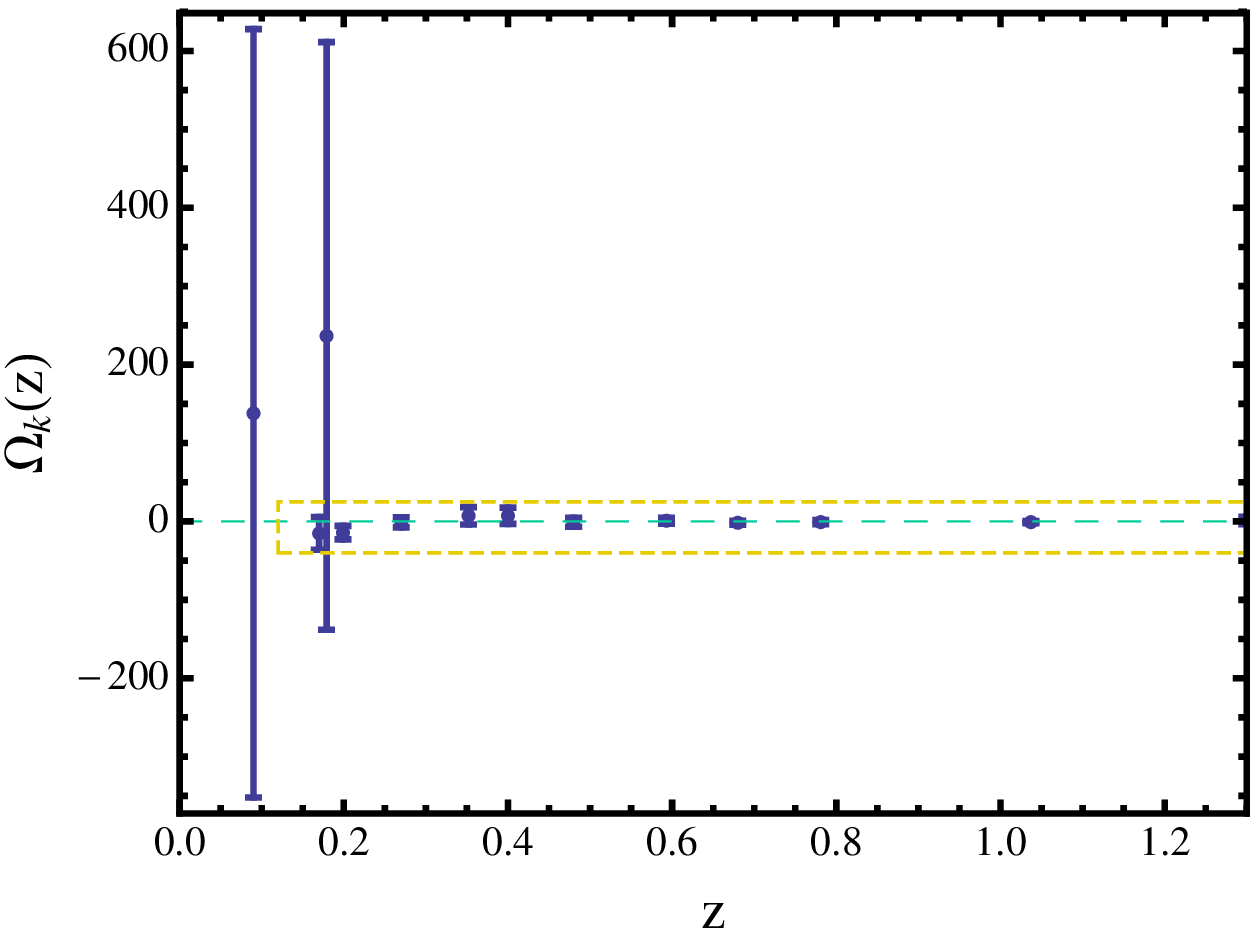}\quad
\includegraphics[scale=0.68]{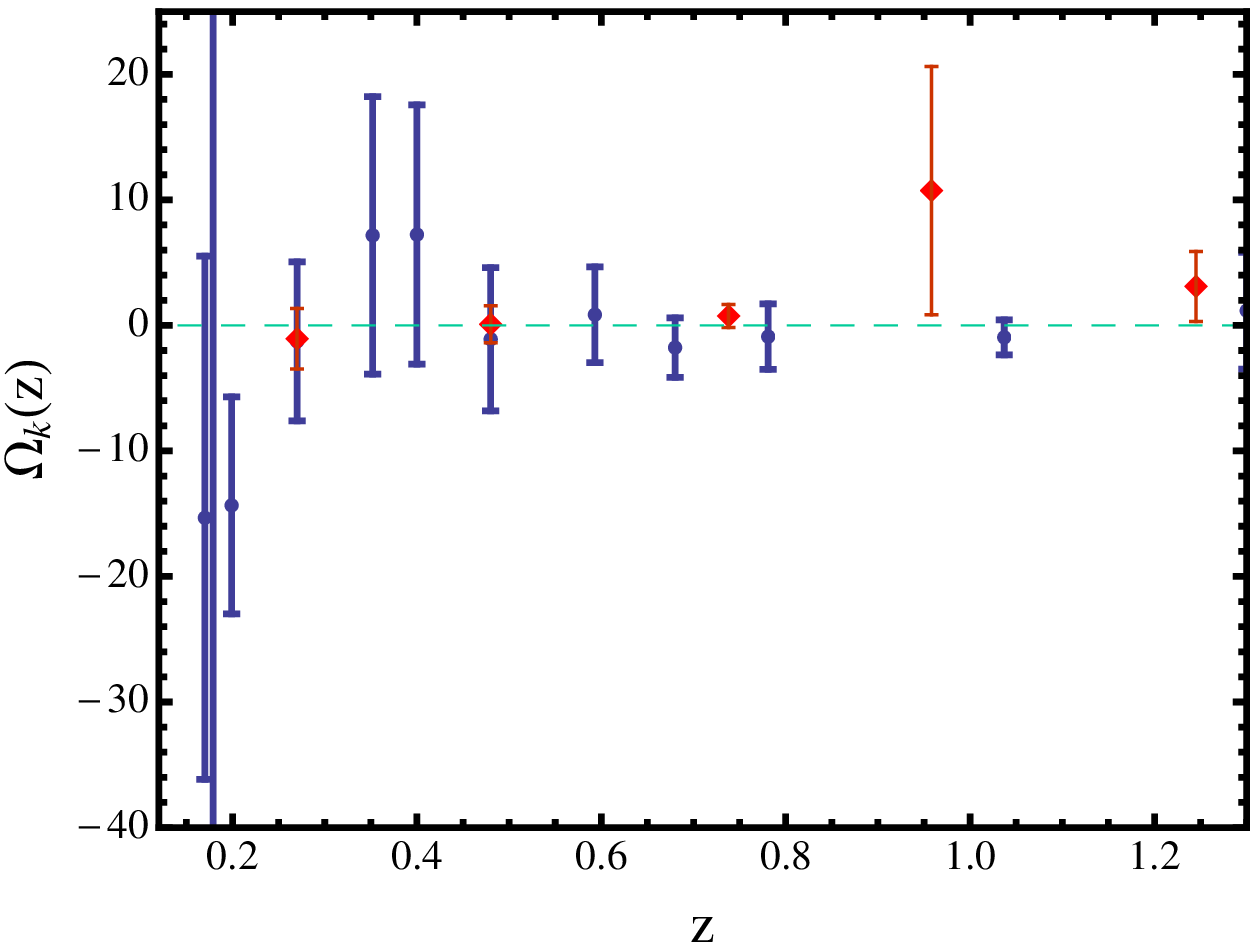}
\caption{The curvature parameter $\Omega_K$ obtained by using  $H(z)$ data from passively evolving galaxies and SnIa data and applying to them the binning procedure described in Sec. \ref{sec:bins1}. For comparison, we also show results from the second binning procedure, described in Sec. \ref{sec:bins2}, as red thin error bars.
The light-blue long-dashed line corresponds to the case of a flat FLRW universe: $\Omega_K(z) = 0$.
On the left panel we show our full results, while on the right panel we focus on the region inside the yellow dashed box, i.e. on redshifts where constraints are best: $z>0.12$. We omit to show horizontal error bars in order not to make the plot too crowded and confusing.}
\label{fig:Omk-bins1}
\end{figure}

The resulting $\Omega_K$ are shown in Fig. \ref{fig:Omk-bins1}.
We notice from the left panel of Fig. \ref{fig:Omk-bins1} that at the smallest and third smallest redshift the error bars are extremely large. Errors reduce noticeably for redshifts $z>0.2$, and, for a better visualisation, on the right panel we focus on the region  inside the yellow dashed box: $0.12<z<1.3$. Here we see that the best constrained $\Omega_K$ ($\sigma_{\Omega_K} = 1.40$) is at $z=1.04$, while at $z=0.2$ the relative error is smallest: $\sigma_{\Omega_K}/|\Omega_K| = 0.60$. However, the smallness of the error at $z = 1.04$ is only apparent and due to the strongly fluctuating $D_i$ close to this redshift, as can be seen from Fig. \ref{fig:comoving-distances}. We will comment more extensively on this in the following section.

A point to understand is the sensitivity to a particular choice of $H_0$. From Eq. (\ref{eq:omegak}) we see that $\Omega_K \propto 1/H_0^2$. When changing e.g. from HST data to the Planck constraint  \cite{Planck} $H_0 = 67.3 \pm 1.2$, this results in a relative change in $\Omega_K$ of 18\%. Since errors on $\Omega_K$ are $\geq 60\%$, and since they depend very little on the error on $H_0$ (errors differ by less than $0.4 \%$ if the error is halved) this does not affect our results. In the future however, it will be important for this test to be effective, that different measurements of $H_0$ give more compatible results.

Since errors are very large when using this binning system, we try a different one in the hope of improving our results.

\subsubsection{Second binning criterion}\label{sec:bins2}

This second criterion aims at maximizing the use of all available data. It uses the full SnIa set, and leaves out only two of the $H(z)$ data. We divide the redshift interval into 6 bins delimited by the following redshifts: $z =\{{0, 0.18, 0.36, 0.6, 0.876, 1.04, 1.45}\}$.
The latter choice was made for two main reasons. First, we want to have at least two $H(z)$ data points in each bin. Second, we do not want to have bins that are too large, in order for the assumption that $D$ and $H(z)$  be constant inside each bin to be still reasonable. Since we need to use the same binning for both datasets, the last two $H(z)$ data points have to be left out again. This is somehow inevitable as the two data-sets have different redshift ranges. We show the properties of the bins described above in Tab. \ref{tab:bins2}.
\begin{table}
\begin{centering}
\begin{tabular}{|c|c|c|c|c|c|c|c|c|c|}
\hline
\textbf{Bins}  & $z_{min}$ &$z_{max}$ & $z_i$ & $\bar z$ &$\#$ of SnIa & $\#$ of H(z) & $\bar D$ & $D_{,z}$ & $\Omega_{K}$ \tabularnewline
\hline \hline
$1$&$0.00$&$0.18$&$0.09$ &$0.057$& $218 $& $3$ & $118.59 \pm 0.68$ & $4113 \pm 44$ & $100 \pm 118$\tabularnewline
\hline
$2$&$0.18$&$0.36$& $0.27$ &$0.27$&$115$& $3$ & $972.3 \pm 6.6$ & $3824 \pm 168$ & $-1.1 \pm 2.4$ \tabularnewline
\hline
$3$&$0.36$& $0.60$&$0.48$&$0.47$&$123$& $3$ & $1741 \pm 15$ & $3008 \pm 406$ & $0.1 \pm 1.5$\tabularnewline
\hline
$4$&$0.60$&$0.88$&$0.74$&$0.73$&$74$&  $3$ &  $2494 \pm 30$ & $3402 \pm 406$ & $0.75 \pm 0.92$\tabularnewline
\hline
$5$&$0.88$& $1.04$&$0.96$&$0.96$&$27$&  $3$ & $3118 \pm 94$ & $6144 \pm 2341$ & $10.7 \pm 9.9$ \tabularnewline
\hline
$6$&$1.04$&$1.45$&$1.25$&$1.23$&$23$&  $2$ & $3804 \pm 94$ & $3359 \pm 1072$ & $3.1 \pm 2.8$ \tabularnewline
\hline
\end{tabular}
\par\end{centering}
\caption{Properties of bins described in Sec. \ref{sec:bins2}  and values of the comoving distance $\bar D$, derivative of the comoving distance $D_{,z}$ and curvature parameter $\Omega_K$ measured in each bin. With $z_{min}$ and $z_{max}$ we indicate the left and right edge of each bin, respectively. $z_i$ refers to the redshift at the centre of each bin, to which the mean distance $\bar D$ has been assigned, while $\bar z$ is the mean redshift of the bin.
\label{tab:bins2}}
\end{table}

We then use the following procedure. First, we compute the weighted average distance, $\bar D$, and weighted average Hubble function, $\bar H$, in each bin. We assign these values to the redshifts at the centre of the bin, i.e. $z_i =\{{0.09, 0.27, 0.48, 0.738, 0.958, 1.245}\}$.
Also here we should in principle assign $\bar{D}$ to the mean redshift $\bar{z}$ of each bin, but here, too, the difference between $\bar{z}$ and $z_i$ is very small (see Tab. \ref{tab:bins2}) except for the first bin, and the error we commit is very small.
 To compute $D_{,z}(z_i)$, we divide the $\Delta z$ of each bin in two equal sub-bins and compute the weighted average distance in each sub-bin, obtaining two distances: $\bar D_{left}$ and $\bar D_{right}$. We assign $\bar D_{left}$ and $\bar D_{right}$ to the average redshift of the corresponding sub-bin, $\bar z_{left}$ and $\bar z_{right}$. We then use Eq. (\ref{eq:der-num}) to compute $D_{,z} (z_i)$. Fig. \ref{fig:H-D-dD} shows as black thick error bars the values of $\bar{H}$, $\bar{D}$ and $D_{,z}$ computed with this binning.

\begin{figure}
\includegraphics[scale=0.68]{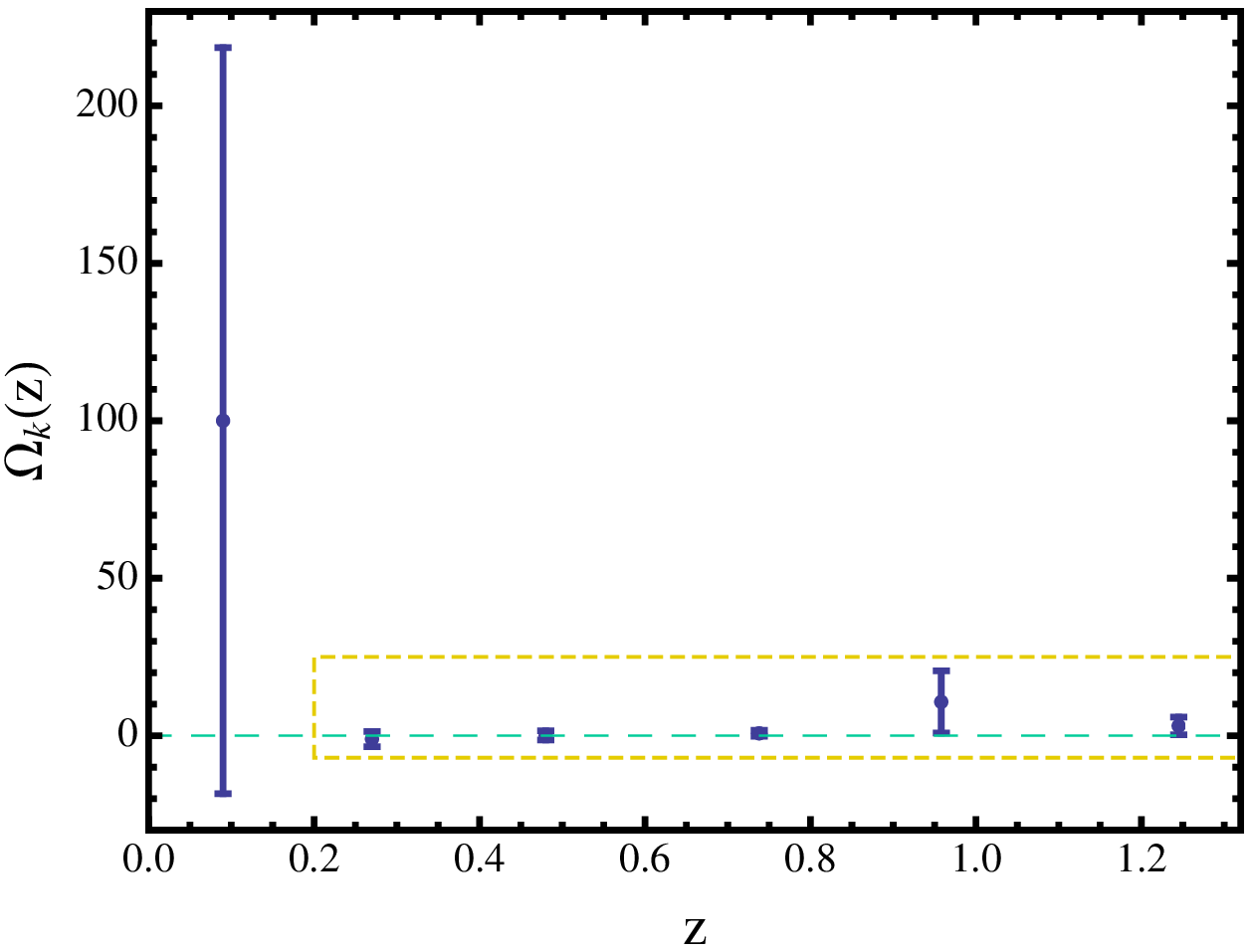}\quad
\includegraphics[scale=0.68]{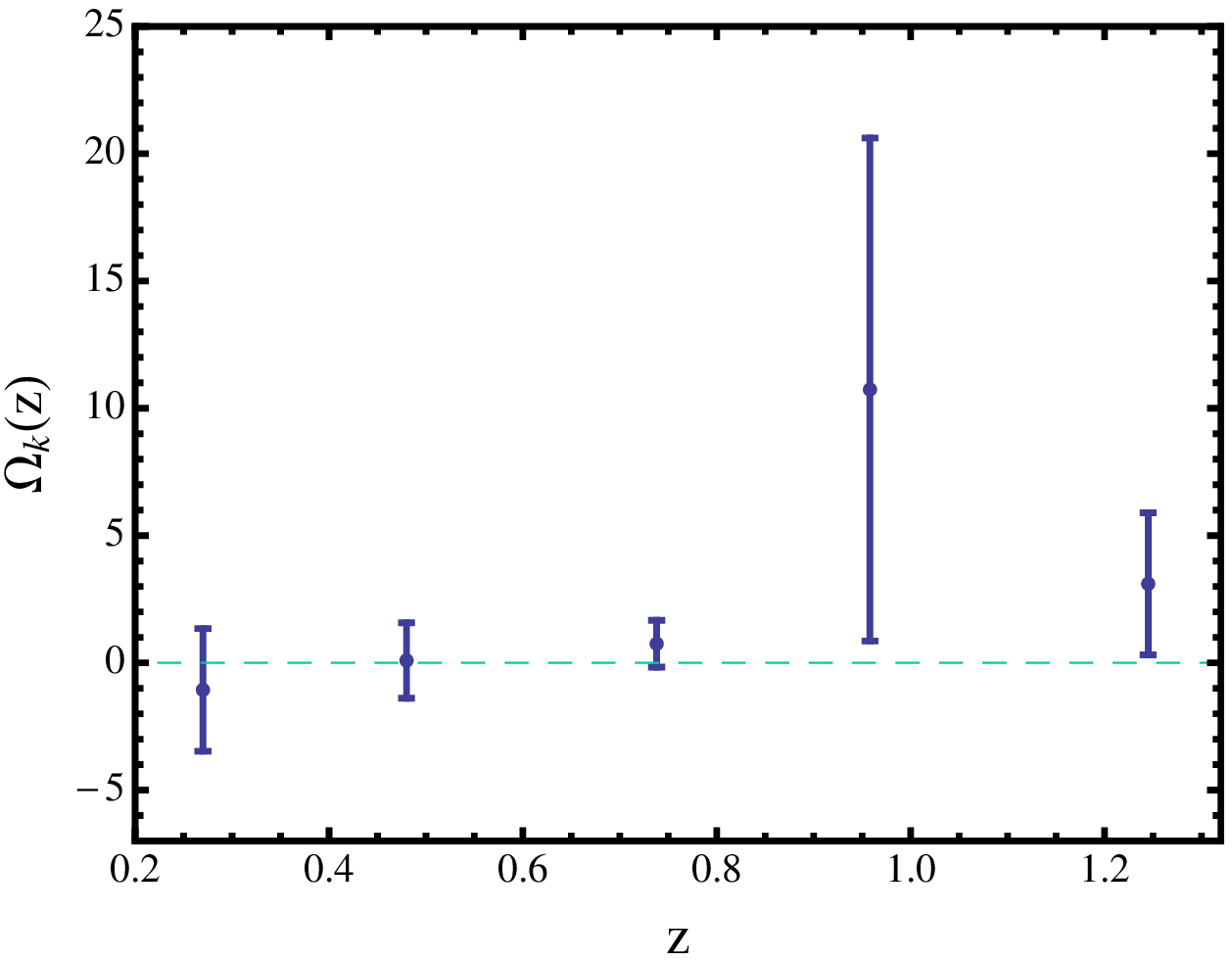}
\caption{The curvature parameter $\Omega_K$ obtained by using  $H(z)$ data from passively evolving galaxies and SnIa data and applying to them the binning procedure described in Sec. \ref{sec:bins2}.
The dashed light-blue long-dashed line corresponds to the case of a flat FRW universe.
On the left panel we show our full results, while on the right panel we focus on the region inside the yellow dashed box, i.e. on redshifts where constraints are best: $z>0.2$. We omit to show horizontal error bars in order not to make the plot too crowded and confusing.}
\label{fig:Omk-bins2}
\end{figure}

We finally compute $\Omega_K(z_i)$ by using Eq. (\ref{eq:omegak}). The results are shown in Fig. \ref{fig:Omk-bins2}.
We notice, by looking at the left panel, that also in this case the error bar of the lowest redshift point is very large. We will see in the following sections that this is a common feature of Eq. (\ref{eq:omegak}), and we will give an explanation for it in the conclusions.
On the right panel we focus on the region corresponding to the yellow dashed box. We can see that the best absolute error ($\sigma_{\Omega_K} =1.48$) is obtained here for $z_i = 0.48$, while the best relative error, $\sigma_{\Omega_K}/|\Omega_K| = 0.9$, is obtained at $z_i = 1.25$.

In order to understand how much the low number of $H(z)$ data affects the size of the error bars, we sum to the measured $H(z)$ data a simulated set with  properties similar to those of the original set.
In particular, the simulated data are built in the same redshift interval, $0.09<z<1.75$, as a random distribution around a Gaussian centred on the best fit flat LCDM cosmology of the $H(z)$ data ($\Omega_m = 0.32$, $H_0 = 0.69$). The scatter around the Gaussian has been taken as the mean error of the real data points, while the error on the simulated data points has been simulated by taking a random distribution around a Gaussian with mean the mean error and variance the variance of the errors of the real data points. In order to reflect the different properties of the data distribution at low and high redshift, the redshift interval has been split into two equal parts, for which the simulated data have been produced separately.
We then analyse the new resulting dataset and find that the errors on $\Omega_K$ remain comparable to those obtained from the smaller real dataset: differences are smaller than 17\%. Even if we simulate a set $10$ times larger than the present one, and if we reduce the scatter and error by a factor of $10$, the difference in $\sigma_{\Omega_K}$, for data at $z>0.2$, is smaller than $64\%$. This means that, in order to obtain a noticeable improvement in the errors, not only the $H(z)$ data but also the SnIa dataset has to improve substantially.

Let us now compare the results of the two different binnings. If we look at the right panel of Fig. \ref{fig:Omk-bins1}, we note that this second binning, corresponding to the red error bars, produces better results everywhere except around $z \sim 1$. After performing a number of tests, we found that this difference is mainly due to the large difference between the values of $D_{,z}$ in the two binnings (see bottom panel of Fig. \ref{fig:H-D-dD}), due in turn to the strong fluctuations of $D$ around $z \sim 1$, that make results in proximity of $z =1$ very binning-dependent.


Although this latter binning produces better results than the previous one, errors on $\Omega_K$ are still very large. In the next section we will therefore describe and use a different technique to measure $\Omega_K$.

\subsection{Principal Component Analysis}\label{sec:pca}

In this section we will use the principal component analysis (PCA) to reconstruct the Hubble parameter and the comoving distance.

Following \cite{nesserisBIA}, the Hubble parameter and the luminosity distance can be modeled in terms of the deceleration parameter $q(z)$. The big advantage of this method, instead of considering the PCA for instance for $w(z)$ as it is common in the literature, is that the results for the deceleration parameter do not depend on $\Omega_m$ or any other parameter. Now, assuming that the deceleration parameter is constant in each bin, let us write it as
\be
q(z) = \sum_{i=1}^{n}q_i\theta(z_i)
\ee
where the $q_i$ are constant in each redshift bin $z_i$ and $\theta(z_i)$ is the step-function,
i.e. $\theta(z_i) = 1$ for $z_{i-1}\leq z \leq z_{i}$ and $0$ elsewhere;
then the Hubble parameter can be written (assuming that $z$ is in the n-th bin) as
\be
H_n(z) = H_0 c_n \left(1+z\right)^{1+q_n}
\label{eq:hubble-pca}
\ee
where the coefficient $c_n$ is
\be
c_n = \prod_{j=1}^{n-1}\left(1+z_j\right)^{q_j-q_{j+1}}\,.
\ee
Consequently, we can evaluate the luminosity distance by simply integrating Eq.~(\ref{eq:hubble-pca}) and we find:
\be
d_{L,n}(z)=\frac{c}{H_0}\left(1+z\right)\left[f_n-\frac{\left(1+z\right)^{-q_n}}{c_n q_n}\right]
\ee
where
\be
f_n =  \frac{\left(1+z_{n-1}\right)^{-q_n}}{c_n q_n}+\sum_{j=1}^{n-1}\frac{\left(1+z_{j-1}\right)^{-q_j}-\left(1+z_{j}\right)^{-q_j}}{c_j q_j}
\ee
and $z_0=0$. Some more details regarding the derivation of the previous equations are
shown in Appendix \ref{PCAdetails}. Also, it should be stressed that in all the calculations related to the PCA, the redshifts $z_i$ correspond to the right edge of the bins and are not the average redshifts of the bins, see also Appendix \ref{PCAdetails}.

We decided to express the above quantities in term of the deceleration parameter $q(z)$; however the reader might think that a natural choice would be to choose a stepwise Hubble parameter. The reason why we choose to parameterize our quantities in terms of $q(z)$  is because the deceleration parameter varies slower at high redshift than, for instance, the Hubble parameter.

Next, we want to find the best fit parameters $q_n$ by using two different data sets:
Hubble measurements and luminosity distance measurements from SnIa. However, the two redshift ranges are different: while the measurements of SnIa reach $z \simeq 1.41$, those of the Hubble parameter reach $z=1.80$. For our purpose, we assume constant $q$ for each redshift bin and we divide the survey into 6 redshift bins up to $z=1.45$; in particular, the binning is the same as that of Sec \ref{sec:bins2}: $\{{0, 0.18, 0.36, 0.6, 0.876, 1.04, 1.45}\}$ (see previous subsection for details). To determine the best fit parameters $q_n$, we use a Monte Carlo Markov Chain (MCMC) method \cite{nesserisweb}; the best fits for both data sets are shown in Tab.~\ref{tab:bestfit-qs}

\begin{table}
\begin{centering}
\begin{tabular}{|c|c|c|c|c|}
\hline
 &\multicolumn{2}{c|}{$H(z)$} & \multicolumn{2}{c|}{SnIa} \tabularnewline
\hline
\textbf{Parameters}  & $q$'s &$1\sigma$ & $q$'s &$1\sigma$ \tabularnewline
\hline
 $q_1$  & $-0.235357$ & $2.1714$ & $-0.516911$&$0.135228$\tabularnewline
\hline
 $q_2$  & $0.125349$ & $0.997989$ & $-0.477117$ & $0.342401$ \tabularnewline
\hline
 $q_3$  & $-0.574964$ & $1.06027$  & $0.290415$& $0.565752$ \tabularnewline
\hline
 $q_4$  & $-0.0301597$ & $0.966473$  &$-0.716238$ &$1.02752$\tabularnewline
\hline
 $q_5$  & $3.03175$ & $1.8193$  &$1.95358$&$3.01709$\tabularnewline
\hline
 $q_6$  &  $-0.301256$ &  $0.905626$  & $-1.94705$& 2.02836\tabularnewline
\hline
\end{tabular}
\par\end{centering}
\caption{Best fit of the $q_n$ parameters from the MCMC simulation and their corresponding $1\sigma$ errors for
both Hubble and SNIa measurements.
\label{tab:bestfit-qs}}
\end{table}

\begin{figure}
\includegraphics[scale=0.7]{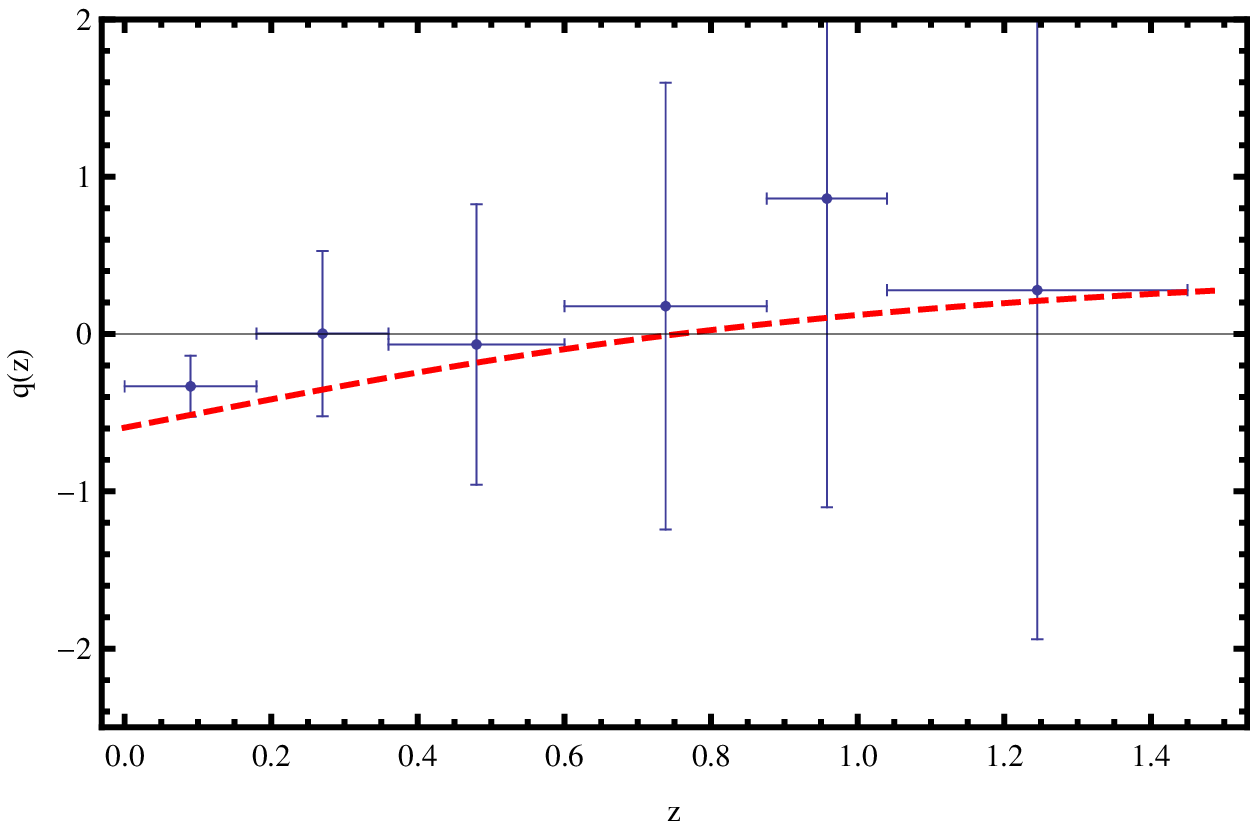}
\includegraphics[scale=0.7]{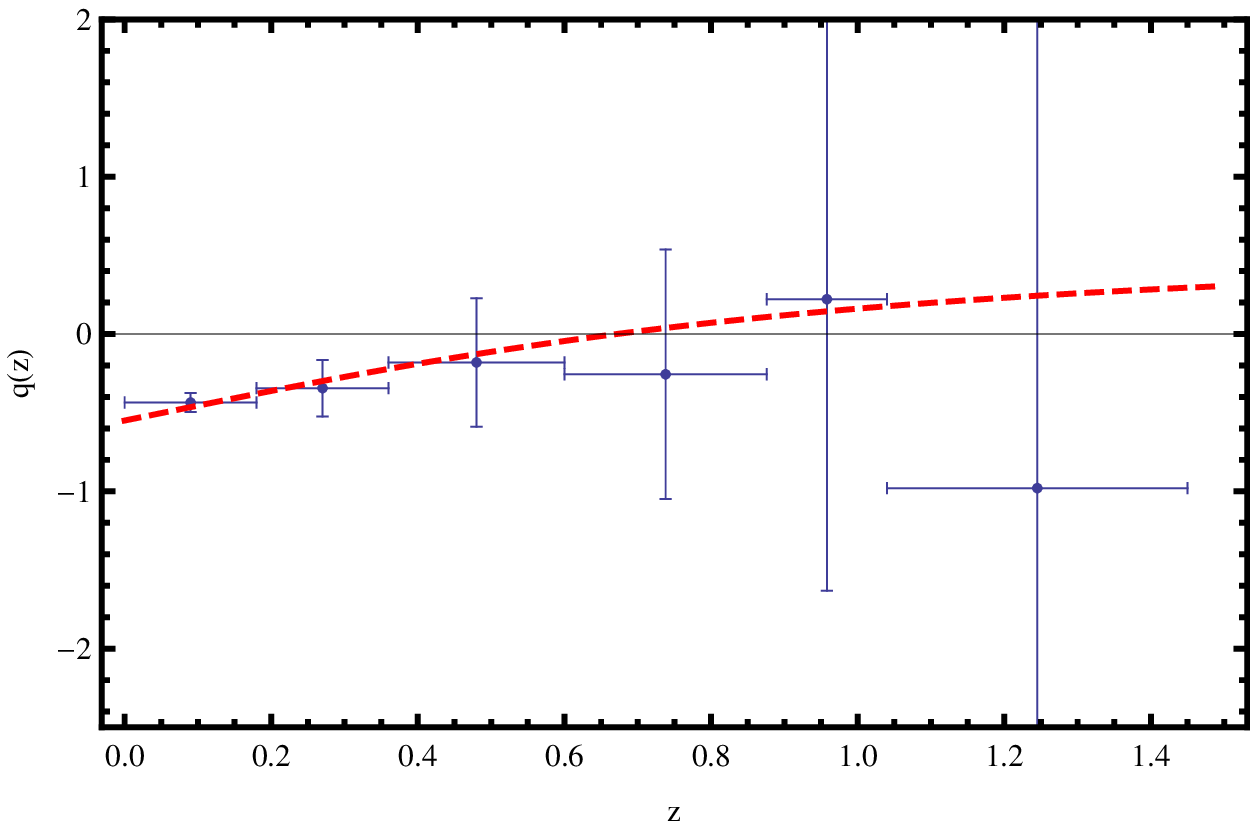}
\caption{We show the deceleration parameter for the two data sets, $H(z)$ (left panel) and SnIa (right panel),
using the PCA technique.
As an example, we also report the deceleration parameter of the $\Lambda$CDM model (red dashed line).}
\label{fig:pca_chain_h_sn}
\end{figure}

Now in order to use the PCA to decorrelate the parameters $q$'s,
we follow Ref.~\cite{hutererch}. We first build a diagonal matrix $\Lambda_{ij}$ with
the eigenvalues of the Fisher matrix $F_{ij}$, which is defined as the inverse of the
covariance matrix $C_{ij}$ (obtained directly from the chains).
Then we define a matrix $\tilde{W}_{ij}=W_{ik}^{T}\,\Lambda_{km}^{1/2}\,W_{mj}$ where
the matrix $W_{km}^{T}$ is the transpose of $W_{km}$ and the latter is a matrix
composed by the eigenvectors of Fisher matrix. We finally normalize $\tilde{W}_{ij}$
such that its rows sum up to unity. The matrix $\tilde{W}_{ij}$ will give the
uncorrelated parameters, i.e. $p_i=\sum_{j=1}^{M}\tilde{W}_{ij}\,q_{j}$, where $M$ is
the total number of parameters. The variance of the parameters $p_i$ will then be
$\sigma^2\left(p_i\right)=1/\lambda_i$. In Fig.~\ref{fig:pca_chain_h_sn} we show the
deceleration parameter $q$ for both $H(z)$ and SnIa measures for our 6 bins.

\begin{table}
\begin{centering}
\begin{tabular}{|c|c|c|c|c|}
\hline
 &\multicolumn{2}{c|}{$H(z)$} & \multicolumn{2}{c|}{SnIa} \tabularnewline
\hline
\textbf{Parameters}  & $q$'s &$1\sigma$ & $q$'s &$1\sigma$ \tabularnewline
\hline
 $q_1$  & $-0.33247$ & $0.19404$ & $-0.435408$ & $0.0609644$ \tabularnewline
\hline
 $q_2$ & $0.00281$ & $0.52534$ & $-0.343985$ & $0.179551$  \tabularnewline
\hline
 $q_3$ & $-0.06639$ & $0.89137$ & $-0.180596$ & $ 0.408267$ \tabularnewline
\hline
 $q_4$& $0.17728$ & $1.42054$ & $-0.256041$ & $0.792702$ \tabularnewline
\hline
 $q_5$ & $0.86084$ & $1.96141$ & $0.221318$ & $1.85372$  \tabularnewline
\hline
 $q_6$ & $0.27898$ & $2.21949$ &  $-0.979667$ &  $3.23438$ \tabularnewline
\hline
\end{tabular}
\par\end{centering}
\caption{Principal component values from the MCMC analysis and their corresponding $1\sigma$ errors for both Hubble and
SNIa measurements.
\label{tab:bestfit-pca}}
\end{table}

Finally, we are in the position to evaluate the values of $\Omega_{K}$ for the six different bins. We want to remind the reader that we are using two different data sets to evaluate the $q$'s parameters: one for $H(z)$ and the other one for $D(z)$. In practice, to evaluate the Hubble parameter we will make use of the best fit of $q$ obtained using the $H(z)$ data, whereas to evaluate the comoving distance $D(z)$ we will use the best fit of $q$ obtained using the SnIa data. The curvature parameter will then be
\be
\Omega_{K}(z, q_n) = \frac{\left[H_n(z;q_{n,H})D(z;q_{n,SN})\right]^2-1}{\left[H_0D(z;q_{n,SN})\right]^2}
\ee
and the errors on $\Omega_K$ have been evaluated by using
\be
\sigma_{\Omega_{K}}^2 = \left.\sum_{i,j}\frac{\partial\Omega_{k, q_n}}{\partial q_{i}}C_{ij}\frac{\partial\Omega_{k, q_n}}{\partial q_{j}}\right|_{H}+\left.\sum_{k,l}\frac{\partial\Omega_{k, q_n}}{\partial q_{k}}C_{kl}\frac{\partial\Omega_{k, q_n}}{\partial q_{l}}\right|_{SN}
\label{eq:errors-omk-pca}
\ee
where $C_{ij}$ and $C_{kl}$ are the covariance matrices of the $q$ parameters using Hubble data and SnIa data, respectively.

\begin{figure}
\includegraphics[scale=0.7]{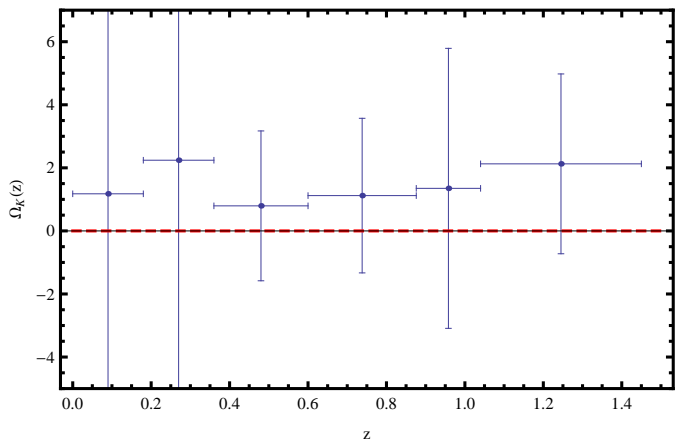}
\caption{We show the curvature parameter computed by combining the two data sets, $H(z)$ and SnIa, using the PCA technique.
The red dashed line refers to the flat FLRW model.}
\label{fig:omegak-pca}
\end{figure}

In Fig.~\ref{fig:omegak-pca} we show the errors on the $\Omega_{k}$ for our 6 bins and
in Tab.~\ref{tab:omegak-pca} we report their values. Clearly, the currently available data are not able
to constrain the curvature parameter with sufficient accuracy and little can be said about
$\Omega_{K}$ as the relative errors are of the order of few $100\%$.

The binning technique, showed in the previous section, and the PCA are somehow equivalent: we decide to divide the survey in different bins and we evaluate all the quantities in each bin. However, as previously stated, the principal component analysis gives substantially better results because we are decorrelating the parameters in each bin. In practice, we use a transformation matrix that is composed by the eigenvectors of the correlation matrix. The eigenvectors form an orthogonal basis to which the old set of parameters are projected to, hence the new parameters (which are a linear combination of the old set of parameters) will be uncorrelated; as a consequence the errors on the new set of parameters will be in general smaller than the errors of the {\em untransformed} parameters.

\begin{table}
\begin{centering}
\begin{tabular}{|c|c|c|}
\hline
\textbf{Parameters}  & $\Omega_K$'s &$1\sigma$  \tabularnewline
\hline
 $\Omega_{K_1}$  & $1.17677$& $25.2965$   \tabularnewline
\hline
 $\Omega_{K_2}$ & $2.24179$ & $7.5252$   \tabularnewline
\hline
 $\Omega_{K_3}$ & $0.794552$ & $2.3758$ \tabularnewline
\hline
 $\Omega_{K_4}$& $1.11986$ & $2.4511$  \tabularnewline
\hline
 $\Omega_{K_5}$ & $1.34971$ & $4.4396$    \tabularnewline
\hline
 $\Omega_{K_6}$ & $2.12658$ & $2.8511$   \tabularnewline
\hline
\end{tabular}
\par\end{centering}
\caption{Values of the curvature parameters at different redshifts and their corresponding $1\sigma$ errors using the PC analysis.
\label{tab:omegak-pca}}
\end{table}

\subsection{Genetic Algorithms}\label{sec:genetic}

In what follows we will briefly describe the Genetic algorithms (GA). For more details and cosmological applications of GA see \cite{Bogdanos:2009ib, Nesseris:2010ep, Nesseris:2012tt}. The GAs are based on the principles of evolution through natural selection, where a group of individuals, the ``population", evolves over time under the joined influence of two operators: the crossover (the combination of two or more different individuals) and the mutation (a random change in an individual). The probability of the ``reproductive success'' that an individual will produce offspring is directly proportional to the fitness of the individual. In our case the fitness is taken to be the $\chi^2$ function, and it measures how accurately each individual describes the data.

The algorithm starts with a group of individuals, the population, which in our case are sets of functions randomly generated based on a predetermined set of functions, the grammar, e.g. $\exp, \sin, \log$ etc, and a chosen set of operations, e.g. $+,-,\times,\div$. In each generation, the fitness for each individual of the population is evaluated and the operations of crossover and mutation are applied. This process is repeated several thousand times until certain termination criteria are reached, e.g. the maximum number of generations or the desired degree of convergence has been achieved. Then the best-fit solution can be used to extract the cosmological parameters or other quantities of interest.

\begin{figure}
\includegraphics[scale=0.5]{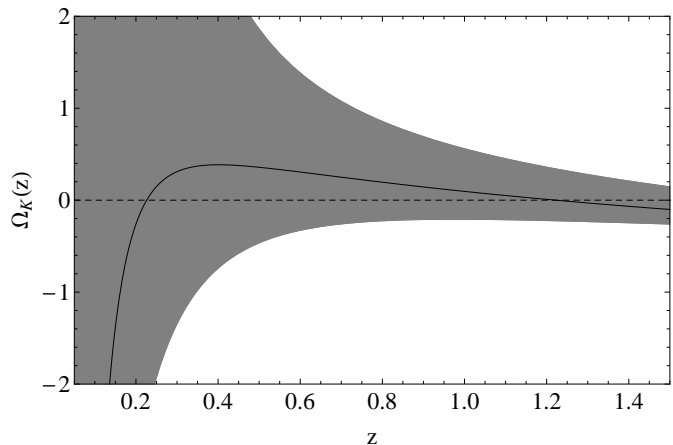}
\caption{The reconstruction of $\Omega_K(z)$ from the GA. In both cases the gray region
is the $1\sigma$ error and the solid line the best-fit. Clearly, it is consistent with $\Omega_K=0$.}
\label{fig:GAplots}
\end{figure}

In Fig. \ref{fig:GAplots} we show the results of the application of the GA on the data and the reconstruction of $\Omega_K$ of Eq. (\ref{eq:omegak}) from the $H(z)$ and SnIa data. In this case, we used the $H(z)$ data to get $H(z)$ directly, while $D(z)$ was obtained from the SnIa. The error regions (gray bands) were made with the path integral formalism of \cite{Nesseris:2012tt} and correspond to the $1\sigma$ error. From Fig. \ref{fig:GAplots} it is clear that the reconstruction is consistent with $\Omega_K=0$. The main advantage of using the GAs in this case is that we can obtain model-independent constraints of the parameters of interest and, as we will see in later sections, they are in excellent agreement especially with the Pad\'e approximation. The reason for the increased errors at small redshift will be explained in the last Section. Overall, we find that the GAs give the smallest error on $\Omega_K$, $\sigma_{\Omega_K}\sim0.1$, due to the fact that the method itself provides a smooth and analytical expression at all redshifts, compared to the binning methods. Also, it is somewhat better than the Pad\'e method due to its better flexibility and non-parametric approach when fitting the data.

\subsection{Pad\'e approximation}\label{sec:pade}

In this section we followed a different approach in reconstructing the Hubble parameter, called {\em Pad\'e approximation}.
The main advantage in following this approach is that it gives the best approximation of a function by using only rational functions.
Following \cite{nesserisBIA} (and references therein), the Hubble parameter and the luminosity distance
can be modeled by only two parameters, $(a,b)$. The $H(z)$ and $D(z)$ are
\be
H(z)=H_0\left[\frac{1+b(1+z)^3}{1+b}\right]^{\frac{a+b}{3b}}
\label{eq:hubble_pade}
\ee
and
\bea
H_0d_{L}(z) = (1+z)(1+b)^{\frac{a+b}{3b}}\left\{(1+z)\,_{2}F_{1}\left[\frac{a+b}{3b},\frac{1}{3},\frac{4}{3};-b(1+z)^3\right]-\,
_{2}F_{1}\left[\frac{a+b}{3b},\frac{1}{3},\frac{4}{3};-b\right]\right\}\,,
\label{eq:dL_pade}
\eea
where $_{2}F_{1}$ is the Gauss hypergeometric function. It is interesting to notice that
we can recover the $\Lambda$CDM model by just setting
$(a,b)= (\Omega_{m_0}/2\Omega_{\Lambda_0},\Omega_{m_0}/\Omega_{\Lambda_0})$
into Eq.~(\ref{eq:hubble_pade}).

Using the $H(z)$ data and SnIa data we can calculate two independent sets of parameters $(a,b)$
(best fit) and their corresponding errors simply by a likelihood analysis.
\begin{itemize}
\item {\bf Hubble data}:

Our $\chi^2$ is defined as usual as
\be
\chi^2 = \sum_{i=1}^{n}\frac{\left[H_i-H(a,b, z_i)\right]^2}{\sigma_i^2}
\ee
where $H_i$ and $\sigma_i$ are the Hubble data and their errors, respectively, and $H(a,b,z_i)$ is given by Eq.~(\ref{eq:hubble_pade}). We marginalize over $H_0$.

The best fit and the corresponding $1\sigma$ errors are: $(a_H,b_H)=(0.23\pm0.07, 0.53 \pm 0.72)$

\item {\bf SnIa data}

In this case, the $\chi^2$ is
\be
\chi^2 = \sum_{i=1}^{n}\frac{\left[\mu_i-\mu\left(a,b, z_i\right)\right]^2}{\sigma_i^2}
\ee
where $\mu_i$ and $\sigma_i$ are the SnIa distance moduli and their errors, respectively;
the distance moludi $\mu(a,b,z_i)$ will be given by Eq.~(\ref{eq:dist-mod}).

The best fit and the corresponding $1\sigma$ errors are: $(a_{SN},b_{SN})=(0.19\pm0.15, 0.39 \pm0.46)$

\end{itemize}

\begin{figure}
\includegraphics[scale=0.7]{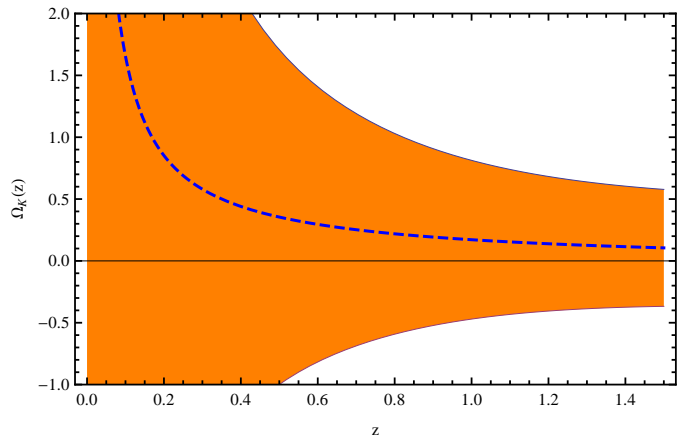}
\caption{We plot the $\Omega_{K}$ as a function of redshift (blue dashed line) and the superposed error region (orange region).}
\label{fig:pade}
\end{figure}

Once the parameters $a$ and $b$ have been found for both datasets, we can reconstruct the Hubble parameter
and comoving distance and hence the curvature parameter. In Fig.~\ref{fig:pade} we plot
the curvature parameter as a function of redshift and the corresponding error region.
The errors have been evaluated using Eq.~(\ref{eq:errors-omk-pca})
where $q_i$ are now simply the two parameters $a$ and $b$.
We find that the constraints that we can give on the curvature parameter are very weak as
the noise is much larger than the constraint itself.

\section{Forecasts from the Euclid Survey}\label{sec:forecasts}

In this section we will now study the sensitivity with which the Euclid experiment\footnote{http://www.euclid-ec.org/}
\cite{Cimatti:2009is, RedBook}  will be able to constrain the curvature parameter $\Omega_{K}$ at different redshifts.

Euclid \cite{RedBook} is a medium-size mission of the ESA Cosmic Vision programme
whose launch is planned for 2020. It will perform two surveys: a photometric survey in the visible and in three near-infrared bands, to measure weak gravitational lensing maps by imaging $\sim 1.5$ billion galaxies, and a spectroscopic slitless survey of $\sim 50$ million galaxies. Both surveys will be able to constrain both the expansion and the growth history of the Universe and will cover a total area of $15,000$ square deg.

Our fiducial Euclid survey follows the specifications which can be found in the
Euclid Definition Study Report (also called Red Book) \cite{RedBook}. As a fiducial model for our Fisher analysis we choose the WMAP-7 flat $\Lambda$CDM cosmology,
as also used in the Euclid  Red Book \cite{RedBook}. This means that we have $\Omega_{m_0} h^2 = 0.13$, $\Omega_{b,0} h^2 = 0.0226$, $\Omega_\Lambda = 1-\Omega_{m_0} =  0.73$, $H_0 = 71 $, $n_s = 0.96$ (where $n_s$ is the scalar spectral index) and $w=-0.95$. The matter power spectrum was computed using CAMB\footnote{http://camb.info} \cite{camb}. More specifically, in this paper we only exploit the BAOs in the galaxy power spectrum as a standard ruler, in order to constrain the angular diameter distance and consequently the curvature parameter.

\subsection{The angular diameter distance}\label{sec:da}

Let us consider again the expression for the curvature parameter:
\be
\Omega_{K}\left(z\right)=\frac{\left[H\left(z\right)D_{,z}\left(z\right)\right]^{2}-1}{\left[H_{0}D\left(z\right)\right]^{2}}\,.
\label{eq:omegak1}
\ee
Our aim in this section is to put constraints on the curvature parameter in different redshift bins starting from BAO data. To this end we need an expression for the angular diameter distance. We can invert Eq.~(\ref{eq:omegak1}) and we have:
\be
\frac{\partial D\lzr}{\partial z}=\pm\frac{1}{H\lzr}\sqrt{1+H_{0}^{2}D^{2}\lzr\Omega_{K}\lzr}\,.
\label{eq:dOmkdz}
\ee
We obtained an expression for the comoving distance which depends
on the curvature density parameter. In order to use the expression above in our Fisher matrix calculation,
we just use the finite approximation of the derivative of a function. Let us rewrite Eq.~(\ref{eq:dOmkdz})
in the following form:
\be
\frac{\partial D\lzr}{\partial z}\simeq\frac{D\left(z_{n+1}\right)-D\left(z_{n}\right)}{\Delta z_{n}}=\pm\frac{1}{H\lzr}\sqrt{1+H_{0}^{2}D^{2}\left(z_{n}\right)\Omega_{K}\left(z_{n}\right)}
\label{eq:dOmk}
\ee
where $\Delta z_{n}$ corresponds to the width of the bins (assumed to be constant). From Eq. (\ref{eq:dOmk}) we can derive the expression for $D\left(z\right)$
valid for each bin:
\be
D_{n+1}=D_{n}+\frac{\Delta z_{n}}{H_{n}}\sqrt{1+\Omega_{K_{n}}H_{0}^{2}\left(1+z_{n}\right)^2D_{n}^{2}}\,.
\label{eq:dOmk-1}
\ee
Since for the BAO Fisher matrix analysis we need the angular diameter distance $D_{A}$, we rewrite  Eq.~(\ref{eq:dOmk-1}) as:
\be
D_{A}(n+1)=\frac{1+z_{n}}{1+z_{n+1}}D_{A}(n)+\frac{1}{1+z_{n+1}}\frac{\Delta z_{n}}{H_{n}}\sqrt{1+\Omega_{K_{n}}H_{0}^{2}\left(1+z_{n}\right)^2D_{A}^{2}(n)}\,.
\label{eq:Da-n}
\ee
where $D_{A}(n)$ refers to the angular diameter distance evaluated in the $n$-th bin.

The derivatives of $D_{A}(n+1)$ with respect to the $\Omega_{K_{j}}$
(curvature parameter in the $j$-th bin) are:
\bea
\frac{\partial D_{A}(n+1)}{\partial\Omega_{K_{j}}} & = & \frac{1+z_{n}}{1+z_{n+1}}\frac{\partial D_{A}(n)}{\partial\Omega_{K_{j}}}+
\frac{\Delta z_{n}}{2H_{n}\left(1+z_{n+1}\right)}\frac{H_{0}^{2}\left(1+z_{n}\right)^2D_{A}^{2}(n)}{\sqrt{1+\Omega_{K_{n}}H_{0}^{2}D_{A}^{2}(n)}}\frac{\partial\Omega_{K_{n}}}{\partial\Omega_{K_{j}}}+\nonumber \\
 & + & \frac{\Delta z_{n}}{H_{n}\left(1+z_{n+1}\right)}\frac{H_{0}^{2}\left(1+z_{n}\right)^2 D_{A}(n)\,\Omega_{K_{n}}}{\sqrt{1+\Omega_{K_{n}}H_{0}^{2}\left(1+z_{n}\right)^2 D_{A}^{2}(n)}}\frac{\partial D_{n}}{\partial\Omega_{K_{j}}}\,.
\label{eq:dDadOmk}
\eea
We choose now to take into account the derivative of the $H_n$
with respect to the curvature parameter $\Omega_{K_{j}}$. This is because we are assuming that
the measurements of the Hubble parameters should be independent of those of $D_A$
(we will come back to this in the next section).
For the Fisher matrix analysis we need to evaluate the derivatives for a reference cosmology.
Our reference cosmology is flat $\Lambda$CDM, hence $\Omega_{K_i}=0$.
For this fiducial model, Eq.~(\ref{eq:dDadOmk}) can be written as:
\be
\frac{\partial D_{A}(n+1)}{\partial\Omega_{K_{j}}}=\frac{1+z_{n}}{1+z_{n+1}}\left.\frac{\partial D_{n}}{\partial\Omega_{K_{j}}}\right|_{\Omega_{K_{j}=0}}+\left.\frac{1}{2H_{n}}\frac{\Delta z_{n}}{1+z_{n+1}}H_{0}^{2}\left(1+z_{n}\right)^2D_{A}^{2}(n)\right|_{\Omega_{K_{j}=0}}\delta_{_{K_{j}K_{n}}}\,.
\label{eq:dDadOmk-ref}
\ee
The second term of the above equation contains a delta of Kronecker,
$\delta_{_{K_{j}K_{n}}}$, that is non zero only if $\Omega_{K_{j}}=\Omega_{K_{n}}$.
The first term on the right hand side of  Eq.~(\ref{eq:dDadOmk-ref}) exists only if $j<n$,
and this can be easily seen from Eq.~(\ref{eq:dOmk-1}), where the
comoving distance in one bin depends on $\Omega_{k}$ of the previous
bin. Let us now assume that $j<n$; then Eq. (\ref{eq:dDadOmk-ref})
becomes:
\be
\frac{\partial D_{n+1}}{\partial\Omega_{K_{j}}}=\frac{1+z_{n}}{1+z_{n+1}}\left.\frac{\partial D_{n}}{\partial\Omega_{K_{j}}}\right|_{\Omega_{K_{j}=0}}\,.
\label{eq:dDadOmk-ref-1}
\ee
This can be further simplified just by realizing that if we take the
derivative of $D_{n}$ with respect to $\Omega_{K_{j}}$ we are still
left with a term containing the delta of Kronecker, which is non zero
only if $j=n-1$, and another term containing the derivative of $D_{n-1}$,
which is non zero only if $j<n-1$. Proceeding by iteration we are
left with:
\bea
\frac{\partial D_{A}(n+1)}{\partial\Omega_{K_{j}}}&=&\frac{1+z_{n}}{1+z_{n+1}}\frac{\partial D_{A}(n)}{\partial\Omega_{K_{j}}}=
\frac{1+z_{n}}{1+z_{n+1}}\frac{1+z_{n-1}}{1+z_{n}}\frac{\partial D_{A}(n-1)}{\partial\Omega_{K_{j}}} = \frac{1+z_{n-1}}{1+z_{n+1}}\frac{1+z_{n-2}}{1+z_{n-1}}\frac{\partial D_{A}(n-2)}{\partial\Omega_{K_{j}}} = \nonumber \\
&=& ... = \frac{1+z_{j+2}}{1+z_{n+1}}\frac{1+z_{j+1}}{1+z_{j+2}}\frac{\partial D_{A}(j+1)}{\partial\Omega_{K_{j}}}
\label{eq:dDadOmk-ref-2}
\eea
hence, Eq.~(\ref{eq:dDadOmk-ref-2}) becomes:
\be
\frac{\partial D_{A}(n+1)}{\partial\Omega_{K_{j}}} =  \frac{1+z_{j+1}}{1+z_{n+1}}\frac{\partial D_{A}(j+1)}{\partial\Omega_{K_{j}}}\,.
\label{eq:dDadOmk-ref-3}
\ee
Let us now assume that $j=n$; then Eq.~(\ref{eq:dDadOmk-ref}) reads:
\be
\frac{\partial D_A(j+1)}{\partial\Omega_{K_{j}}}=\left.\frac{\left(1+z_{j}\right)^2}{2H_j}\frac{H_{0}^{2}D_A^{2}(j)\Delta z_{j}}{1+z_{j+1}}\right|_{\Omega_{K_{j}=0}}.\label{eq:dDadOmk-sec}
\ee
We remind the reader that Eqs.~(\ref{eq:dDadOmk-ref-2})
and (\ref{eq:dDadOmk-sec}) are valid as long as all the $\Omega_{K_{j}}$
in the reference cosmology are zero (of course a more general formula
could be found but this goes beyond the goal of this work).
We have now two different derivatives for the angular diameter distance depending on the bin.
By iteration, we can write Eq.~(\ref{eq:Da-n}), in the reference cosmology, as:
\be
D_{A}(n+1)=\frac{1+z_{1}}{1+z_{n+1}}D_{A}(1)+\frac{1}{1+z_{n+1}}\sum_{k=1}^{n}\frac{\Delta z_{n}}{H_{n}}\,.
\label{eq:Da-n1}
\ee
Then, Eqs.~(\ref{eq:dDadOmk-ref-2}) and (\ref{eq:dDadOmk-sec}), together with Eq.~(\ref{eq:Da-n1}), become:
\be
\frac{\partial D_{A}\left(n+1\right)}{\partial\Omega_{K_{j}}}=
\begin{cases}
\left.\frac{H_{0}^{2}}{2H_j}\frac{\Delta z_{j}}{1+z_{j+1}}\left[\left(1+z_1\right)D_{A}(1)+\sum_{k=1}^{j-1}{\frac{\Delta z}{H_k}}\right]^2\right|_{\Omega_{K_{j}=0}}, & \mbox{if } j=n \\
\left.\frac{H_{0}^{2}}{2H_j}\frac{\Delta z_{j}}{1+z_{n+1}}\left[\left(1+z_1\right)D_{A}(1)+\sum_{k=1}^{j-1}{\frac{\Delta z}{H_k}}\right]^2\right|_{\Omega_{K_{j}=0}}, & \mbox{if } j<n
\end{cases}
\ee
The derivatives of the logarithm of the angular diameter distance with respect to the curvature parameter,
needed for the computation of the BAO Fisher matrix are finally:
\be
\frac{\partial \ln D_{A}\left(n+1\right)}{\partial\Omega_{K_{j}}}=
\begin{cases}
\left.\frac{H_{0}^{2}}{2H_j}\Delta z_{j}\frac{\left[\left(1+z_1\right)D_{A}(1)+\sum_{k=1}^{j-1}{\frac{\Delta z}{H_k}}\right]^2}{\left(1+z_1\right)D_{A}(1)+\sum_{k=1}^{j}{\frac{\Delta z}{H_k}}}\right|_{\Omega_{K_{j}=0}}, & \mbox{if } j=n \\
\left.\frac{H_{0}^{2}}{2H_j}\Delta z_{j}\frac{\left[\left(1+z_1\right)D_{A}(1)+\sum_{k=1}^{j-1}{\frac{\Delta z}{H_k}}\right]^2}{\left(1+z_1\right)D_{A}(1)+\sum_{k=1}^{n}{\frac{\Delta z}{H_k}}}\right|_{\Omega_{K_{j}=0}}, & \mbox{if } j<n
\end{cases}
\label{eq:Da-fisher}
\ee

\subsection{Fisher matrix formalism}\label{sec:fisher}

Let us now show and comment on the Fisher matrix forecasts for the
Euclid galaxy \emph{redshift} survey. Following \cite{se}, we write the observed galaxy power spectrum as:
\bea
P_{obs}(z,k_{r}) & = & \frac{D_{Ar}^{2}(z)H(z)}{D_{A}^{2}(z)H_{r}(z)}G^{2}(z)b(z)^{2}\left(1+\beta\mu^{2}\right)^{2}P_{0r}(k)\nonumber \\
&  & +P_{shot}(z)
\eea
where the subscript $r$ refers to the values assumed for the reference
cosmological model, i.e. the model at which we will evaluate the Fisher
matrix. Here $P_{shot}$ is the shot noise due to discreteness in
the survey, $\mu$ is the cosine of the angle of the wave mode with respect to the line of sight, $P_{0r}(k)$ is the present matter power spectrum for the fiducial cosmology,
$G(z)$ is the linear growth factor of matter perturbations,
$b(z)$ is the bias factor. The wavenumber $k$ has also to be written in terms of the fiducial cosmology (\cite{se} and see also \cite{aqg} and \cite{sa} for more details).

The spectroscopic survey covers a redshift range of $0.65<z<2.05$, which we divide into 14 bins of equal width $\Delta z = 0.1$. Regarding the bias, we assume it to be scale-independent, since this is a quite good approximation for the large linear scales which we will use. Our fiducial bias was derived by \cite{Orsi}  using a semi-analytical model of galaxy formation, and it is the same bias function used for the Euclid Red Book forecasts. The expected galaxy number densities which we used can be found in \cite{myeuclid} and were computed by using a sophisticated simulation \cite{Garillietal}. The maximum scale $R$ used are such that $\sigma^2(R)\leq0.25$, with an additional cut at $k_{\rm max}= 0.20 \,h\,{\rm Mpc}^{-1}$ to avoid non-linearity problems. The wavenumber $k$ is also to be transformed between the fiducial cosmology and the general one.

The parameters that we use for evaluating the Fisher matrix are shown
in Tab. \ref{tab:Cosmological-parameters}.
We evaluated the linear matter power spectrum $P_{0r}$ using CAMB, see \cite{camb}. Once we have the full Fisher matrix, we marginalize over all the parameters but the angular diameter distances.
We obtain a submatrix with only $D_{A}$, ${\cal F}_{mn}$. The new Fisher matrix for the
$\Omega_{K}$ will be given by:
\be
F_{_{\Omega_{K_{i}}\Omega_{K_{j}}}}=\frac{\partial\log D_{A}(m)}{\partial\Omega_{K_{i}}}{\cal F}_{mn}\frac{\partial\log D_{A}(n)}{\partial\Omega_{K_{j}}}
\ee
where the derivatives $\frac{\partial\log D_{A}(n)}{\partial\Omega_{K_{j}}}$
are given by Eq.~(\ref{eq:Da-fisher}).
In this work we are using only the information coming from the measurements of the angular diameter distance and
we are ignoring the information of the Hubble parameter (by simply neglecting the derivative of Hubble parameter
with respect to the $\Omega_K$'s).
This is because the Fisher matrix analysis has been done assuming a reference cosmology, i.e. flat $\Lambda$CDM;
with this assumption, the Hubble parameter and the angular diameter distance are not independent quantities,
so taking into account in our analysis the two parameters would bias our results.
Choosing to neglect the derivatives of the Hubble parameter with respect to $\Omega_K$ means
that we assume $H(z)$ to be measured with infinite precision by another experiment.
Anticipating the results, our analysis shows that even with this strong, and very optimistic, assumption the sensitivity
of the Euclid satellite in constraining the curvature parameter is not enough to
rule out inhomogenous models, like for instance the LTB model shown in Fig.~(\ref{fig:errors-euclid-SN}).
In \cite{lar}, the authors also forecasted the measurement of the curvature parameter with the Euclid survey,
using a different fiducial cosmology.
In their analysis they consider both the Hubble parameter and angular diameter distance;
this is why the errors they found on the curvature parameter are of a factor $2-3$ larger than our results.

\begin{table}
\begin{centering}
\begin{tabular}{|c|c|c|}
\hline
 & \textbf{Parameters}  & \tabularnewline
\hline
1  & Reduced Hubble  & $h$\tabularnewline
\hline
2  & Total matter density  & $\omega_{m}=\Omega_{m_{0}}h^{2}$\tabularnewline
\hline
3  & Total baryon density  & $\omega_{b}=\Omega_{b_{0}}h^{2}$\tabularnewline
\hline
4  & Spectral index  & $n_{s}$\tabularnewline
\hline
5  & Growth index  & $\gamma$\tabularnewline
\hline
 &  & \tabularnewline
\hline
 & \emph{For each redshift bin}  & \tabularnewline
\hline
 &  & \tabularnewline
\hline
6  & Shot noise  & $P_{s}$\tabularnewline
\hline
7  & Redshift space distortion  & $\log\beta$\tabularnewline
\hline
8  & Growth Factor  & $\log G$\tabularnewline
\hline
9  & Angular diameter distance  & $\log D_{A}$\tabularnewline
\hline
10  & Hubble  & $\log H$\tabularnewline
\hline
\end{tabular}
\par\end{centering}

\caption{Cosmological parameters for the Fisher matrix analysis
\label{tab:Cosmological-parameters}}
\end{table}

\subsection{Adding the supernovae}\label{sec:addingSN}

To add the SnIa, we first compute the corresponding Fisher Matrix.
We present the details of the cumbersome calculations in Appendix \ref{sniafisher}.
In this case, the most important quantity is the marginalized error on $\log D_{i}/D_{1}$, which is given by \cite{lucashinjibook}
\be
(F)_{ii}^{-1}=\frac{2\sigma^{2}}{a^{2}N}\approx0.42\frac{\sigma^{2}}{N}.
\ee

\subsection{Results\label{results}}

\begin{figure}
\includegraphics[scale=0.7]{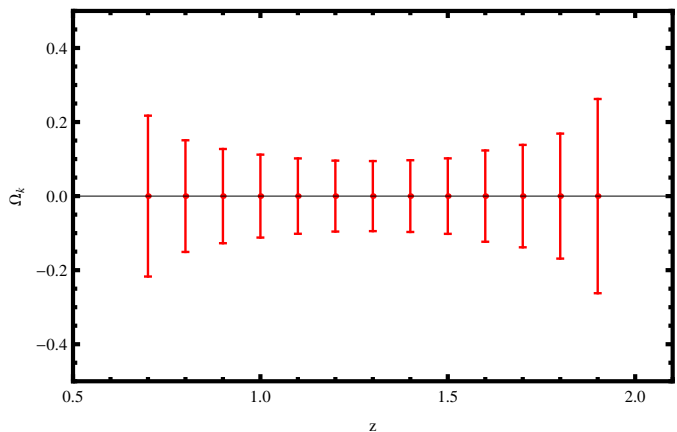}
\caption{Error bars for the curvature parameter using the Euclid survey only.}
\label{fig:errors-euclid}
\end{figure}

\begin{figure}
\includegraphics[scale=0.7]{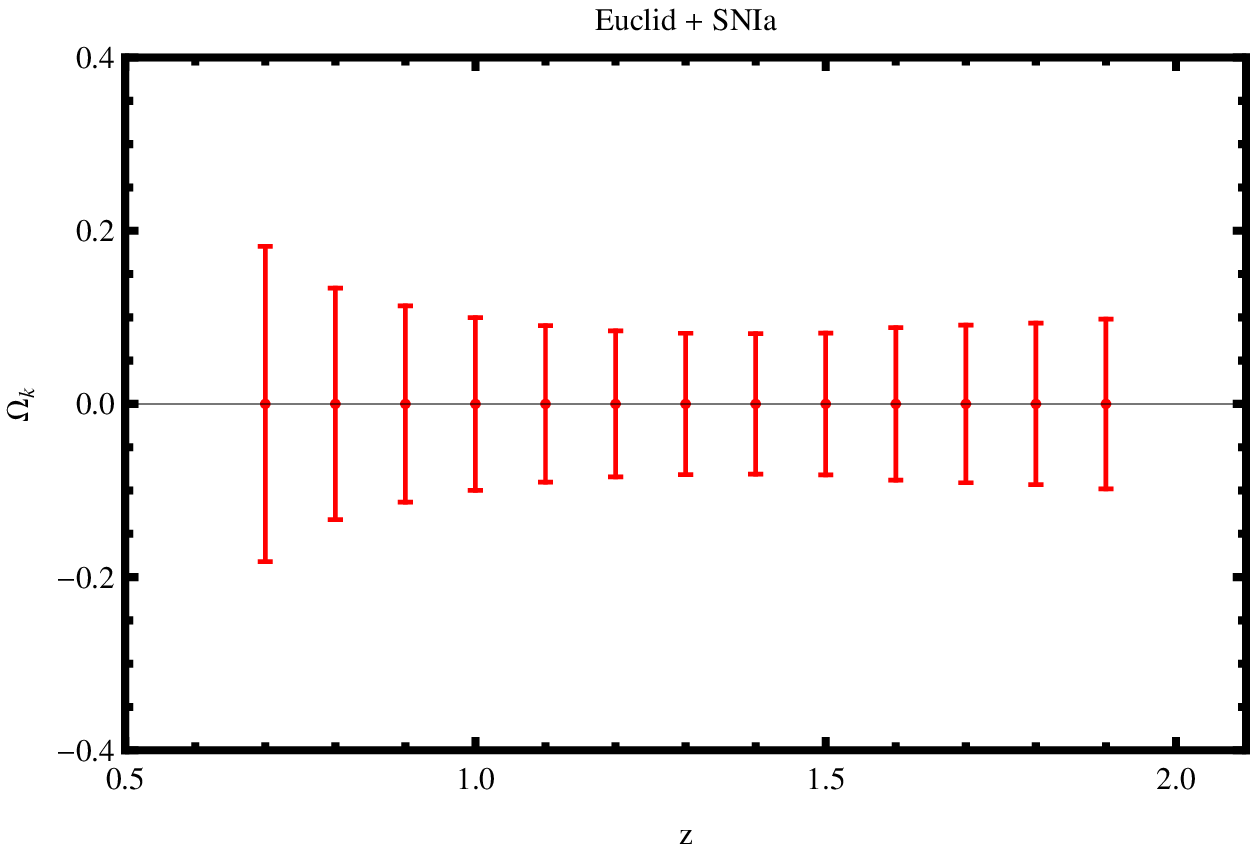}
\includegraphics[scale=0.7]{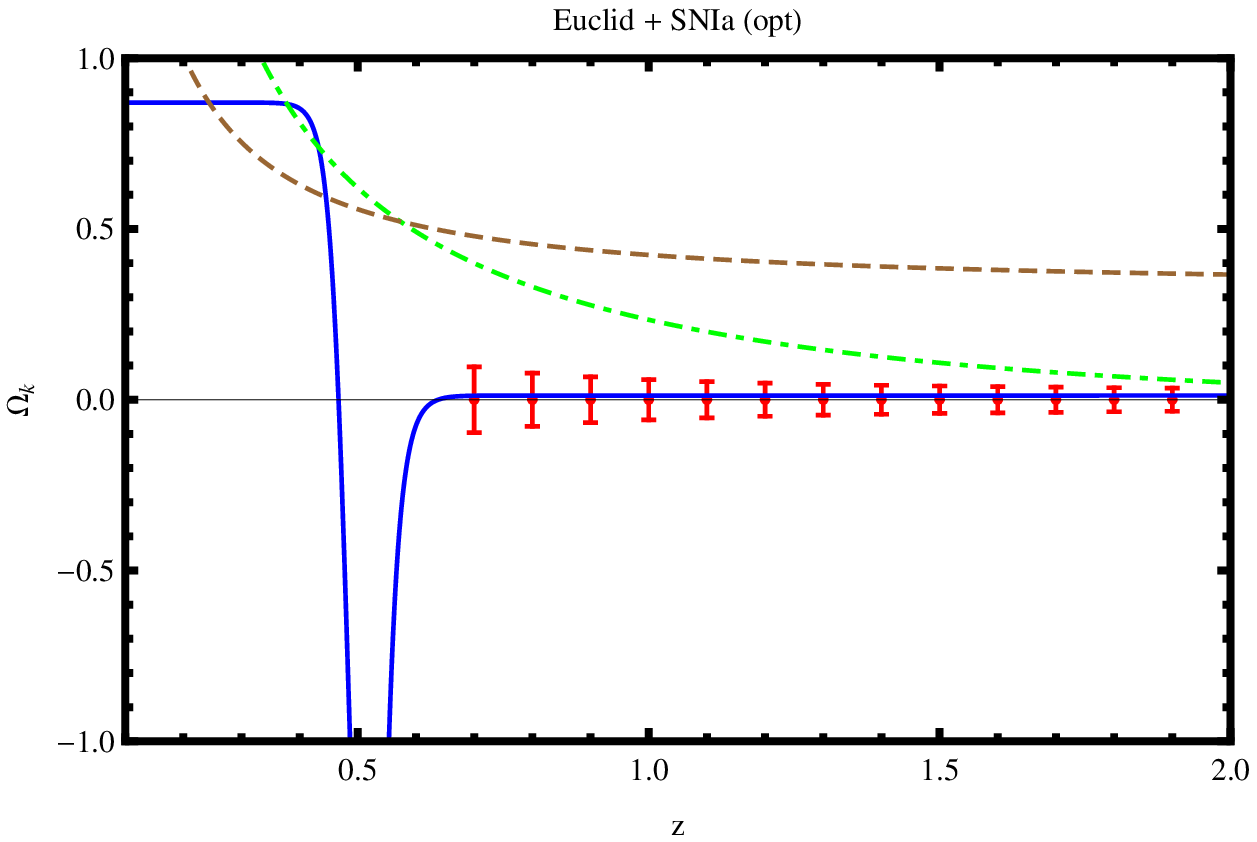}
\caption{Error bars for the curvature parameter using Euclid and adding SnIa with: N = 100 and N=1000, left and right respectively.
In the right panel we also plotted, as an example, three non-FLRW models. The blue solid line corresponds to the curvature parameter in the LTB model.  The brown dashed line indicates the Tardis model as in Fig. 12 a) of \cite{s3} and is courtesy of  Mikko Lavinto and Syksy R{\"a}s{\"a}nen. The green dot-dashed line corresponds to the timescape scenario, see \cite{wil, Duley:2013ksw} and is courtesy of David Wiltshire. }
\label{fig:errors-euclid-SN}
\end{figure}

In Fig.~\ref{fig:errors-euclid} we show the errors on $\Omega_{K}$ for each redshift bin.
As we can see, we loose the information on the last bin; this is due to the fact that the information on the curvature parameter
in one bin depends on the angular diameter distance of the next bin.
In Fig.~\ref{fig:errors-euclid-SN} we show the errors on $\Omega_{K}$ for each redshift bin when adding to the Euclid forecasts the SnIa for two cases, i.e.
for a number of SnIa N = 100 and for N = 1000. In Tab.~\ref{tab:euclid-errors} we report the errors for each bin for Euclid only and for Euclid with the addition of the SnIa.

\begin{table}
\begin{centering}
\begin{tabular}{|c|c|c|c|c|c|c|c|c|c|c|c|c|c|c|}
\hline
&$z$  & $0.7$  & $0.8$  & $0.9$  & $1.0$  & $1.1$  & $1.2$  & $1.3$& $1.4$ & $1.5$ & $1.6$ & $1.7$ & $1.8$ & $1.9$   \tabularnewline
\hline
{\rm Euclid}&$\sigma\left(\Omega_{k}\right)$  & $0.2172$  & $0.1508$  & $0.1272$  & $0.1119$  & $0.1017$  & $0.0957$  & $0.0946$ & $0.0968$ & $0.1019$ & $0.1232$ & $0.1382$ & $0.1687$ & $0.2621$ \tabularnewline
\hline
Case 1&$\sigma\left(\Omega_{k}\right)$  & $0.1820$  & $0.1336$  & $0.1133$  & $0.0998$  & $0.0903$  & $0.0843$  & $0.0816$ & $0.0811$ & $0.0819$ & $0.0881$ & $0.0910$ & $0.0932$ & $0.0981$ \tabularnewline
\hline
Case 2&$\sigma\left(\Omega_{k}\right)$  & $0.0962$  & $0.0781$  & $0.0671$  & $0.0591$  & $0.0530$  & $0.0484$  & $0.0450$ & $0.0424$ & $0.0401$ & $0.0386$ & $0.0369$ & $0.0352$ & $0.0338$ \tabularnewline
\hline
\end{tabular}
\par\end{centering}
\caption{Errors for the curvature parameter using Euclid only, adding SnIa case 1) and case 2).
\label{tab:euclid-errors}}
\end{table}

\footnotetext{courtesy of David Wiltshire.}

We notice that using Euclid data alone, the redshift region where errors are smallest is $1.1 < z < 1.5$.
At lower and at higher redshifts constraints become worse by up to a factor of two and a half.
Adding the first set of $N=100$ SnIa does not change results substantially at low redshift,
while it does improve constraints at redshifts $z > 1.5$. The second set of SnIa with $N=1000$
instead improves constraints at all redshifts by a factor of $\sim 2$ and it moves the best
constrained area to $ z >1.5$. In this case, one will have to worry about modeling the SnIa systematic errors,
but this goes beyond the scope of our paper and is left to future work.

For comparison, we also plotted in the right panel of Fig.~\ref{fig:errors-euclid-SN} the curve representing the behavior of the
curvature parameter for the timescape scenario of \cite{wil}, for the Tardis cosmology of \cite{s3}, and for the LTB model given by \cite{GBH}.
The Euclid survey will not be able to rule out the latter class of models, not even when adding SnIa constraints, as these models are asymptotically flat homogeneous models. 
Only surveys that observe galaxies at low redshifts, with $z<0.5$, will be able to rule out (or confirm)
these models since they all manifest, at these redshifts, a very different behavior from the flat FLRW model,
as can be seen from Fig.~\ref{fig:errors-euclid-SN}.
However, if a model has a value of $\Omega_K$ of about $\sim 0.4$, we may be able to distinguish it from flat FLRW with Euclid future data.

\section{Conclusions \label{sec:conclusions}}

In this paper we used four different methods to reconstruct, in a model-independent fashion, the Hubble parameter and the luminosity distance, measured by two different data sets. We find that all four methods used, i.e. direct binning, principal component, genetic algorithms and Pad\'e approximation, give results that are in agreement with respect to their general behavior.

In particular, all four methods lead to very large errors at low redshifts, while errors decrease noticeably for larger $z$. The transition redshift is in all cases between $0.2$ and $0.4$. To be specific, as for the direct binning and the PCA techniques, the best measured $\Omega_K$ have errors $\sigma_{\Omega_K} \simeq 3$.
For the genetic algorithms and Pad\'e approximations instead, we found that the errors on the curvature parameter are  of the order of about $\sigma_{\Omega_K}\sim 20$ at small redshifts, and they decrease when $z$ grows, reaching a value of about $\sigma_{\Omega_K}\sim 0.5$ for redshifts $z\geq1$. This means that reconstruction techniques manage to reduce the error if compared to techniques where data need to be binned. The best method, leading to the smallest error on $\Omega_K$, $\sigma_{\Omega_K}\simeq 0.1$, is the one based on GA.

We also find that the choice of the binning, as expected, influences the values of the errors, by affecting mostly the derivative of the comoving distance.
This is simply due to the nature of the SnIa data: in regions where the values of the distance modulus have large fluctuations, there will be a strong dependence of the choice of binning on the value of the derivative.

Going back to the problem of the large size of the errors at low redshifts, this may appear strange because here we have a large number of data with small errors (in particular as regards SnIa data, see e.g. Fig. \ref{fig:comoving-distances}). If we have a closer look though, we find that the main reason for this feature lays on the expression of $\Omega_K$ itself, so let us explain this in detail.
The curvature parameter was found to be
\be
\Omega_{K} = \frac{\left[H(z)D_{,z}(z)\right]^2-1}{\left[H_0D(z)\right]^2}\,.
\label{eq:omegak-concl}
\ee
This expression tells us that if the Universe is not homogeneous then the curvature parameter has to vary with redshift, i.e. $\Omega_K$ is not constant.
From Eq.~(\ref{eq:curvAngDist}) we learn that the comoving distance is the integral of the Hubble parameter; however, Eq.~(\ref{eq:curvAngDist}) has been evaluated under the assumption of homogeneity, as a consequence $\Omega_K$ can only be constant.

Now let us assume that, via one observable, we measure the Hubble parameter $H(z)$ and, via another independent observable, we measure the luminosity distance, which gives a slightly different Hubble parameter, say $H_1(z)$. Then, substituting $H_1(z)$ in Eq.~(\ref{eq:curvAngDist}), Eq.~(\ref{eq:omegak-concl}) will look like:
\be
\tilde{\Omega}_{K}(z) = \Omega_{K}\frac{1 -\left(\frac{H(z)}{H_1(z)}\right)^2\cos^2\left[\sqrt{-\Omega_K}\int_{0}^{z}{\frac{H_0{\rm d}x}{H_1(x)}}\right]}{\sin^2\left[\sqrt{-\Omega_K}\int_{0}^{z}{\frac{H_0{\rm d}x}{H_1(x)}}\right]}\,,
\label{eq:omegak-lowz}
\ee
where $\tilde{\Omega}_K$ is the function that we want to evaluate, whereas $\Omega_K$ is the standard curvature parameter. As we can see, for small redshifts, the numerator of Eq.~(\ref{eq:omegak-lowz}) stays finite ($\cos(0) =1$), but the denominator does not ($\sin(0)=0$). Let us assume that for $z\rightarrow 0$, the quantity $\sqrt{\Omega_K}\int_{0}^{z}{H_0{\rm d}x/H_1(x)}\rightarrow \epsilon$, where
$\epsilon$ is a small number that goes to zero when $z = 0$. Then, Eq.~(\ref{eq:omegak-lowz}) reads
\be
\tilde{\Omega}_{K}(z) = \Omega_{K}\frac{1-\left(\frac{H(z)}{H_1(z)}\right)^2\cos^2\left(\epsilon\right)}{\sin^2\left(\epsilon\right)}=
\Omega_{K}\frac{1-\left(\frac{H(z)}{H_1(z)}\right)^2+\left(\frac{H(z)}{H_1(z)}\right)^2\sin^2\left(\epsilon\right)}{\sin^2\left(\epsilon\right)}
\label{eq:omegak-lowz1}
\ee
 For $\epsilon \rightarrow 0$, Eq.~(\ref{eq:omegak-lowz1}) becomes
\be
\tilde{\Omega}_{K}(z)=\lim_{\epsilon \rightarrow 0}\left(\Omega_{K}\frac{1-\left(\frac{H(z)}{H_1(z)}\right)^2+\left(\frac{H(z)}{H_1(z)}\right)^2\sin^2\left(\epsilon\right)}{\sin^2\left(\epsilon\right)}
\right) \rightarrow \infty\,,
\ee
Then, in order for the above equation to stay finite at small redshift we need $H_1(z)$ to be exactly equal to $H(z)$; in other words, the curvature parameter stays constant at small redshifts if the Universe is homogeneous! This is the reason why for small redshifts, when measuring $H(z)$ and $D(z)$ with independent observations, the
values of the curvature parameter (and their errors) are very large. In this paper, we used two different datasets which give, of course, two different best fits of the parameters and consequently these two best fits will give two different Hubble parameters ($H(z)$ and $H_1(z)$), so that Eq.~(\ref{eq:omegak-lowz1}) still diverges when $z\rightarrow 0$.

Alternatively, we can also perform a series expansion on Eq.~(\ref{eq:omegak-concl}) for small $z$. Doing so we find
\be
\tilde{\Omega}_{K}=\Omega_k+\frac{1-\frac{H_{1,0}^2}{H_0^2}}{z^2}+\frac{\frac{2 H_0 H'(0)-H_{1,0} H_1'(0)}{H_0^2}-\frac{H_1'(0)}{H_{1,0}}}{z}+...,
\ee
where the primes denote a derivative with respect to $z$, eg $H_1'(0)=\frac{dH_1}{dz}|_{z=0}$, while $H_0$ and $H_{1,0}$ are the values of the two Hubble parameters at $z=0$. The advantage of this approach is that we can also clearly see, in a model-independent way and without assuming a Dark Energy model, the type of the singularity as $z\rightarrow0$. So, unless $H(z)=H_1(z)$ then there will be a singularity $\sim1/z^2$ at $z=0$.

To also prove numerically what is explained above, we report the case of the PCA analysis. Here the value of the Hubble parameter  in each of the 6 bins, obtained when using the best fits from SnIa are: $H(z_i) =\{71.1431, 76.9306, 87.9404, 103.691, 108.447, 129.132\}$. When using the best fits from $H(z)$ measurements from passively evolving galaxies, we obtain instead: $H(z_i) =\{71.1061, 82.0685, 91.8856, 102.916, 131.697, 166.138\}$. The difference of the Hubble parameters in the first redshift bin is $0.037$; even though this number is very small, it is not small enough to guarantee the curvature parameter to be stable at low redshifts.

The same conclusion was previously found by \cite{s3}\footnote{However, here a divergence $\propto z^{-1}$ was found}. 
Also \cite{sha} explain this feature of the $\Omega_K$ test, but their analysis was done assuming a particular dark energy 
model and the results could be different for different models. Here we have demonstrated that  Eq.~(\ref{eq:omegak-concl}) 
always diverges at small redshifts independently of the model as $\sim1/z^2$, for all models different from 
the homogeneous and isotropic one.

As a last remark, it is important to notice that in general the data are not free of systematics (and they are unavoidable), 
so even if the universe is homogeneous and isotropic, we will always reconstruct two different Hubble parameters from 
two different datasets, which will lead to a divergent (or very large) $\Omega_K$ at low redshifts. 
Probably the test of $\Omega_K$ will give a results that it is still consistent with the FLRW universe within the errors. Also, one should consider that there will always be a ``noise'' in the measurement of $\Omega_K$ due to the presence of inhomogeneities, that generate fluctuations on the measured values of the pure FLRW cosmological parameters \cite{wessel}.

Future data from the Euclid satellite will improve considerably the errors on the curvature parameter with respect to present data; in particular we will have an improvement of about $10$ times when using Euclid only, and even up to $40$ times if we add future SnIa data to the Euclid survey data.

\section*{Acknowledgments}
We thank Luca Amendola who participated in the beginning of this project and for useful and interesting discussions afterwards.
We also thanks Licia Verde and Antonio Enea Romano for fruitful discussions.
S.~N. and D.~S. acknowledge financial support from the Madrid Regional
Government (CAM) under the program HEPHACOS S2009/ESP-1473-02, from MICINN
under grant AYA2009-13936-C06-06 and Consolider-Ingenio 2010 PAU (CSD2007-00060),
as well as from the European Union Marie Curie Initial Training Network UNILHC
Granto No. PITN-GA-2009-237920. E.~M. was supported by the Spanish MICINNs Juan de la Cierva programme (JCI-2010-08112), by CICYT through the project FPA-2012-31880, by the Madrid Regional
Government (CAM) through the project HEPHACOS S2009/ESP-1473 under grant P-ESP-00346 and by the European Union FP7 ITN INVISIBLES (Marie Curie Actions, PITN- GA-2011- 289442). All authors also acknowledge the support of the Spanish MINECO's ``Centro de Excelencia Severo Ochoa" Programme under Grant No. SEV-2012-0249.

\appendix
\section{The PCA \label{PCAdetails}}

The expressions for the PCA in terms of the deceleration parameter were derived in
Ref. \cite{nesserisBIA}, but here we will present some more details about their derivation.
Let us write the deceleration parameter as
\be
q(z) = \sum_{i=1}^{n}q_i\theta(z_i),
\ee
where $q_i$ are constant in each redshift bin $z_i$ and $\theta(z_i)$ is the theta-function, i.e.
$\theta(z_i) = 1$ for $z_{i-1}\leq z \leq z_{i}$ and $0$ elsewhere.
The general expression for the deceleration parameter is
\be
1+q(z)=\frac{d \ln(H(z))}{d\ln(1+z)},
\ee
and this can be rewritten as
\be
\ln(H(z)/H_0)=\int_0^z\frac{1+q(x)}{1+x}dx,
\ee
or
\bea
H(z)/H_0&=&e^{I(z)}\\
I(z)&=&\int_0^z\frac{1+q(x)}{1+x}dx.
\eea
For $z \in(z_{i-1},z_i]$ and using the fact that $q$ is constant in each bin, we can break the integral $I(z)$ in parts as
\bea
I(z)&=&\int_0^{z_1}(...)+\int_{z_1}^{z_2}(...)+...+\int_{z_{i-1}}^{z}(...)\nn\\
&=&(1+q_1)\ln(1+x)|_0^{z_1}+(1+q_2)\ln(1+x)|_{z_1}^{z_2}+...+(1+q_i)\ln(1+x)|_{z_{i-1}}^{z}\nn\\
&=&(1+q_1)\ln(1+z_1)+(1+q_2)\ln\left(\frac{1+z_2}{1+z_1}\right)+...+(1+q_i)\ln\left(\frac{1+z}{1+z_{i-1}}\right).
\eea
Grouping the constant terms, the Hubble parameter can then be written, for $z$ in the n-th bin, as
\be
H_n(z) = H_0 c_n \left(1+z\right)^{1+q_n}
\ee
where the coefficient $c_n$ is
\be
c_n = \prod_{j=1}^{n-1}\left(1+z_j\right)^{q_j-q_{j+1}}\,.
\ee
We can now follow a similar procedure to calculate the luminosity distance.
Using the definition of the luminosity distance along with the previous equations we have
\bea
d_L(z)&=&\frac{c}{H_0} (1+z) \int_0^z\frac{1}{H(z)/H_0} dz \nn \\
&=&\frac{c}{H_0} (1+z)\left( \int_0^{z_1}(...)+\int_{z_1}^{z_2}(...)+...+\int_{z_{i-1}}^{z}(...)\right)\nn\\
&=&\frac{c}{H_0} (1+z) \left( \frac{1-(1+z_1)^{-q1}}{c_1 q_1}+\frac{(1+z_1)^{-q_2}-(1+z_2)^{-q_2}}{c_2 q_2}+...+ \frac{(1+z_{i-1})^{-q_i}-(1+z)^{-q_i}}{c_{i} q_i}\right).
\eea
Collecting the constant terms, the latter can be written as
\be
d_{L,n}(z)=\frac{c}{H_0}\left(1+z\right)\left[f_n-\frac{\left(1+z\right)^{-q_n}}{c_n q_n}\right],
\ee
where
\be
f_n =  \frac{\left(1+z_{n-1}\right)^{-q_n}}{c_n q_n}+\sum_{j=1}^{n-1}\frac{\left(1+z_{j-1}\right)^{-q_j}-\left(1+z_{j}\right)^{-q_j}}{c_j q_j}\,,
\ee
and $z_0=0$.

\begin{figure}
\includegraphics[scale=0.6]{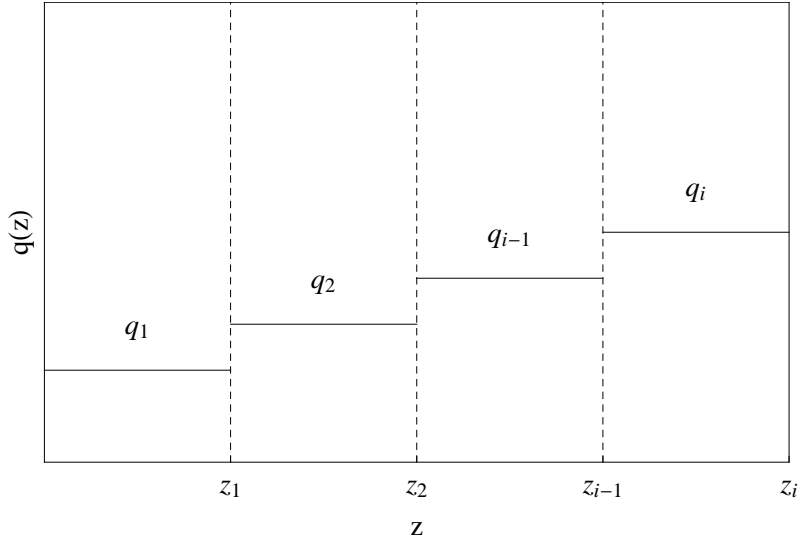}
\caption{Example of a binning scheme. The derivations assume that the point of interest has $z \in(z_{i-1},z_i]$, so that $q(z)=q_i$.}
\label{fig:Cij}
\end{figure}

\section{The SnIa Fisher Matrix \label{sniafisher}}
The SnIa likelihood is
\be
\mathcal{L}=-2\log L=\sum\frac{(m_{i}-m_{t,i}+M)^{2}}{\sigma_{i}^{2}}
\ee
where $M$ is the overall offset (sum of the SnIa absolute magnitude,
the Hubble constant and other things). We marginalize over $M$
\be
L=\int\exp[-\frac{1}{2}\sum\frac{(m_{i}-m_{t,i}+M)^{2}}{\sigma_{i}^{2}}dM=Ne^{-\frac{1}{2}[S_{2}-\frac{S_{1}^{2}}{S_{0}}]}
\ee
where
\be
S_{m}=\sum\frac{(m_{i}-m_{t,i})^{m}}{\sigma_{i}^{2}}
\ee
To find the Fisher Matrix (FM) we calculate
\be
F_{ij}=\frac{\partial(-\log L)}{\partial p_{i}\partial p_{j}}=\frac{1}{2}[S_{,ij}^{2}-\frac{2}{S^{0}}(S_{,i}^{1}S_{,j}^{1}+S^{1}S_{,ij}^{1})]
\ee
where the cosmological parameters are the luminosity distances: $p_{i}=D_{i}$
inside each of the $b$ bins. We have
\be
m_{ti,j}=a\delta_{ij},\quad a=5/\log10\approx2.17 \nonumber
\ee
where $\delta_{ij}$ is unity if the $i$-th SnIa belongs to the
$j$-th bin, 0 otherwise. Then we have
\bea
S_{,i}^{1} & = & -aS^{0B_{i}}\\
S_{,ij}^{1} & = & 0\\
S_{,ij}^{2} & = & 2a^{2}S^{0B_{i}}\delta_{ij}
\eea
where
\be
S^{0B_{i}}=\sum_{B_{i}}\frac{1}{\sigma_{i}^{2}} \nonumber
\ee
Then\be
F_{ij}=2a^{2}\left[S^{0B_{i}}\delta_{ij}-\frac{S^{0B_{i}}S^{0B_{j}}}{S^{0}}\right]
\ee
Suppose now that all errors are identical and equal to $\sigma$, so that $S^{0}=n/\sigma^{2},S^{0B_{i}}=N_{B_{i}}/\sigma^{2},\sum N_{B_{i}}=n$.
Then
\be
F_{ij}=2a^{2}\frac{N_{B_{i}}}{\sigma^{2}}\left[\delta_{ij}-\frac{N_{B_{j}}}{n}\right]
\ee
This matrix is singular because
\be
\sum_{i}\left(N_{B_{i}}\delta_{ij}-\frac{N_{B_{i}}N_{B_{j}}}{n}\right)=N_{B_{j}}-N_{B_{j}}=0\nonumber
\ee
We can however fix the first bin, i.e. assume that we know $D_{1}$ and
then use as parameters $\log D_{i}/D_{1}$. The whole procedure remains
the same because $\log D_{1}$ just adds to $M$ and we just have
to strip off the FM by the first row and column. We can put $N_{B_{i}}=N$
for simplicity, and having $b$ bins, $bN=n$. Then we get
\be
F=a^{2}\frac{N}{\sigma^{2}}\left[I-\frac{1}{b}U\right]
\ee
where $I$ is the identity matrix and $U$ is formed by 1's everywhere.
We can use this matrix to model the SnIa Fisher matrix. For instance
if we have $b=$3 bins, with $N$ SnIa each, then $F$ is a 2x2 matrix
for $\log D_{2}/D_{1}$ and $\log D_{3}/D_{1}$:
\be
\frac{Na^{2}}{\sigma^{2}}\left(\begin{array}{cc}
1-\frac{1}{3} & -\frac{1}{3}\\
-\frac{1}{3} & 1-\frac{1}{3}\end{array}\right) \nonumber
\ee
If needed, we must put zeros for rows and columns for unconstrained
parameters (e.g. $D_{1},H_{i}$). We assume two cases: 1) current data and
2) future data. For case 1) we put $\sigma=0.3$ and $N=100$, while
for case 2) we assume the same $\sigma$ but $N=1000$.

The inverse of a matrix of rank $r$
\be
M=\beta(I-\alpha U)
\ee
is
\be
M^{-1}=\beta^{-1}\left(I-\frac{\alpha}{r\alpha-1}U\right)
\ee
Our FM is of this form, with $r=b-1$, $\alpha=1/b$ and $\beta=a^{2}N/\sigma^{2}$.
Hence
\be
F^{-1}=\frac{\sigma^{2}}{a^{2}N}\left[I+U\right].
\ee
The marginalized error on $\log D_{i}/D_{1}$ is finally
\be
(F)_{ii}^{-1}=\frac{2\sigma^{2}}{a^{2}N}\approx0.42\frac{\sigma^{2}}{N}.
\ee

{}

\end{document}